\documentclass[notitlepage]{article}

\usepackage{natbib}
\usepackage{titlesec}
\usepackage[titletoc,toc,title]{appendix}
\titleformat{\subsection}{\normalfont\itshape}{\thesubsection.}{0.5em}{}
\titleformat{\subsubsection}{\normalfont\itshape}{\thesubsubsection.}{0.6em}{}
\usepackage{adjustbox}
\usepackage{booktabs,siunitx,rotating}
\usepackage{lineno}

\usepackage{enumitem}
\usepackage{pgf,tikz}
\usepackage{appendix}
\usepackage{mathrsfs}
\usepackage{amsmath}
\usepackage{multirow}
\usepackage{booktabs} 
\newcommand{\ra}[1]{\renewcommand{\arraystretch}{#1}}
\usepackage{mathtools}
\usepackage{array}
\usetikzlibrary{arrows}
\usepackage{pdflscape}
\usepackage{textcmds}
\usepackage{amssymb}
\usepackage{hyperref}
\usepackage{textcmds}
\usepackage{caption}
\usepackage[nottoc,notlot,notlof]{tocbibind}
\usepackage{etoolbox}
\usepackage{lipsum} 
\usepackage[utf8]{inputenc} 
\usepackage[T1]{fontenc}    
\usepackage{hyperref}       
\usepackage{url}            
\usepackage{booktabs}       
\usepackage{amsfonts}       
\usepackage{nicefrac}       
\usepackage{microtype}      
\usepackage{sidecap}
\usepackage{lscape}
\usepackage{wrapfig}
\usepackage{multirow}
\usepackage{graphicx}
\usepackage{dcolumn}
\usepackage{subcaption}
\usepackage{indentfirst}
\usepackage{amsmath, amssymb, mathtools, bm, amsthm}
\usepackage{amsfonts}
\usepackage{gensymb}
\usepackage{placeins}
\usepackage{booktabs} 
\usepackage{todonotes}
\usepackage{hyperref}
\usepackage[ruled,vlined]{algorithm2e}
\usepackage{algorithmic}
\usepackage{bm}
\usepackage{cleveref}

\usepackage{pifont}
\title{House Price Determinants and Market Segmentation in Boulder, Colorado:\\ A  Hedonic Price Approach
}

\author{\stepcounter{footnote} Mahdieh Yazdani\thanks{\noindent Department of Economics, Colorado University at Boulder. Email contact: mahdieh.yazdani@colorado.edu}}

\begin{document}

\begingroup
\maketitle
\thispagestyle{empty}

\begin{abstract}
 
In this research we perform hedonic regression model to examine the residential property price determinants in the city of Boulder in the state of Colorado, USA. The urban housing markets are too compounded to be considered as homogeneous markets. The heterogeneity of an urban property market requires creation of market segmentation. To test whether residential properties in the real estate market in the city of Boulder are analyzed and predicted in the disaggregate level or at an aggregate level we stratify the housing market based on both property types and location and estimate separate hedonic price models for each submarket. The results indicate that the implicit values of the property characteristics are not identical across property types and locations in the city of Boulder and market segmentation exists.\\

\noindent \textit{keywords}: Urban housing market, Real estate market appraisal, House price indexes.\\

\end{abstract}
\thispagestyle{empty}
\newpage
\tableofcontents
\thispagestyle{empty}
\endgroup  

\clearpage
\setcounter{page}{1}
\section{Introduction}

House is one of the essentials for human beings to rest and gather with family. In most families dwelling is one of the most important components of a household's wealth (see e.g., \cite{arvanitidis2014economics}). Consequently, to make informed decisions the need for efficient and accurate real estate market price prediction models is apparent.\\

The price of any residential property is affected by many property factors such as its structural, neighbourhood, and spatial attributes. To predict future dwelling valuations the development of accurate housing valuation and property price prediction models is essential (see e.g., \cite{fan2018house}). Hedonic price models have been a common approach to analyze and predict real estate assets. A hedonic method is trying to explain the value of a composite good by decomposing its attributes and estimating their marginal contribution or implicit valuations (see e.g., \cite{rosen1974hedonic}). The hedonic price approach has been extensively employed in many different fields of research, and these studies are often used by policy makers and economists for the analysis of differentiated goods like housing whose individual features do not have explicit market prices.\\

\cite{rosen1974hedonic} used the hedonic regression models in the field of real estate and urban economics. Thereafter, residential hedonic analysis has become widely used as an assessment tool to model real estate markets (see e.g.,  \cite{blomquist1981hedonic}, \cite{milon1984hedonic}, \cite{haase2013tools}, \cite{jiang2014new}, \cite{del2017hedonic}, \cite{waltl2018estimating}, and \cite{krol2020application}). The residential hedonic price approach has been utilized in the property market literature to investigate the relationship between house prices and housing attributes. The primary interests for such an application are improving the precision of house price indexes and providing methods for property value estimates (see e.g., \cite{sheppard1999hedonic}). \\

In this paper we follow the hedonic price approach to develop property appraisal models in the city of Boulder, Colorado. In the city of Boulder, a popular mountain city, with nearly $16,701$ inhabitants, buying a house is considered as one of the most profitable investments and most people in Boulder know the benefit of owning a house. \\

Development of a dwelling price prediction method can significantly improve the efficiency of the housing market. This study brings about some general implications for actual and potential homeowners, property developers, investors, tax assessors, appraisers, brokers, mortgage lenders, banks, and policy makers (see e.g., \cite{frew2003estimating}). The contributions of this paper are as follows:\\

\begin{itemize} [leftmargin=*]
 \item  Collecting real estate datasets from different resources consisting of Multiple Listing Service (MLS) database, Public School Ratings, Colorado Crime Rates and Statistics Information, CrimeReports, WalkScore, and US Census Bureau. 
\end{itemize} 

\begin{itemize} [leftmargin=*]

 \item Screening the collected data set, data cleansing, and applying imputation and winsorization methods. 
\end{itemize} 

\begin{itemize} [leftmargin=*]

 \item Detecting multicollinearity problem.
\end{itemize} 

\begin{itemize} [leftmargin=*]
 \item Applying the hedonic price approach in the aggregate market level to develop property appraisal models and using White’s standard errors test in the presence of heteroskedasticity.
\end{itemize}

\begin{itemize} [leftmargin=*]
 \item Stratifying the housing market in the city of Boulder based on both property types and location to create a number of homogeneous submarkets and performing the hedonic models in different submarket levels.

\end{itemize}

The rest of the paper is organized as follows. Section 2 provides an overview of the literature that employs hedonic regression models. Following this, section 3 explains the empirical models, section 4 presents the data set and the variables, and section 5 elaborates the methodologies used in this paper. The results obtained from performing hedonic regression methods described in section 6. Finally, conclusions and implications are provided in section 7.\\

\section{Related Literature}

Hedonic price analysis originates from \cite{waugh1928quality}, \cite{at1939hedonic}, \cite{stone1954measurement}, \cite{adelman1961index}, and \cite{lancaster1966new}. \cite{lancaster1966new} states that \qq{the good, per se, does not give utility to the consumer; it possesses characteristics, and these characteristics give rise to utility}. For example a house can be viewed as an aggregation of individual attributes such as (lot size, number of bedrooms, bathrooms, and garages) which are implicitly embodied in the house and its utility is generated by characteristics of the house. \\

The hedonic regression model has numerous applications. Hedonic approaches have been used to construct quality-adjusted price indexes for differentiated goods. Namely, hedonic price model has been applied to estimate agricultural commodities value (see e.g., \cite{ethridge1982hedonic} and \cite{wilson1984hedonic}). Several studies have verified the ability of the hedonic regression model to explain the property rental prices determinants (see e.g., \cite{zhang2017quantile}, and \cite{gluszak2018development}). Hedonic regression models have been used to explain variations in dwelling prices and to determine the impact of specific attributes on house prices. Many studies explored the influence of distance to the central business district (CBD), employment centers, park, shopping center, or school on housing values using hedonic housing price models. Examples of these literature include \cite{malpezzi1981flight}, \cite{adair2000house}, \cite{soderberg2001estimating}, \cite{chin2006influence}, \cite{ottensmann2008urban}, \cite{cao2012hedonic}, and \cite{tong2020hedonic}. All found an overall declining pattern of property values with distance from these locations. In addition, a number of studies have applied the hedonic price technique to quantify the effects of environmental features like air, river, and noise pollution on house values (see e.g., \cite{anselin2008errors}, \cite{mei2020valuing}, \cite{chen2017environmental},  \cite{chang2013hedonic}, and \cite{bishop2020best}. \\

\cite{butler1980cross} argues that the implicit assumption in hedonic price methods is that all potential house buyers and suppliers interact in a single market and the market can be characterised by a single hedonic equation determined by the interaction of a supply and demand in a unitary market. However, the urban housing markets are too compounded to be considered as homogeneous goods. Some studies examined heteroscedasticity in hedonic house price models (see e.g., \cite{fletcher2000heteroscedasticity}, \cite{stevenson2004new}, and \cite{kestens2006heterogeneity}). The evidence indicated the presence of heteroscedasticity. \\

The heterogeneity of an urban property market gives rise to its likely segmentation. Real estate market is spatial in nature and many property markets may exist simultaneously within a single urban area. \cite{day2003submarket} states the property market is not a global or even a national market. It is argued that to obtain unbiased estimates of the implicit prices identifying segmentation of the property market is essential. Some studies, namely, \cite{adair1996hedonic}, \cite{watkins2001definition}, \cite{berry2003estimation}, \cite{lipscomb2005household}, \cite{tu2007spatial}, \cite{liu2020modeling}, and \cite{nishi2021illusion} utilized hedonic price techniques to test for market segmentation. They recognised the importance of both spatial and structural characteristics when defining submarkets and observed greater levels of homogeneity in the sample at a submarket level. They conclude that dividing the market into the submarkets normally improves the predictive power of the hedonic regression models. \\

In this study we stratify the housing market in the city of Boulder into a number of distinct submarkets to create a number of homogeneous submarkets from the aggregate heterogeneous dwelling market based on both property types and location.\\

\section{Hedonic Price Model}

\begin{figure}[h!]
\centering
\hspace*{.00001\textwidth} 
\includegraphics[width=0.95\textwidth]{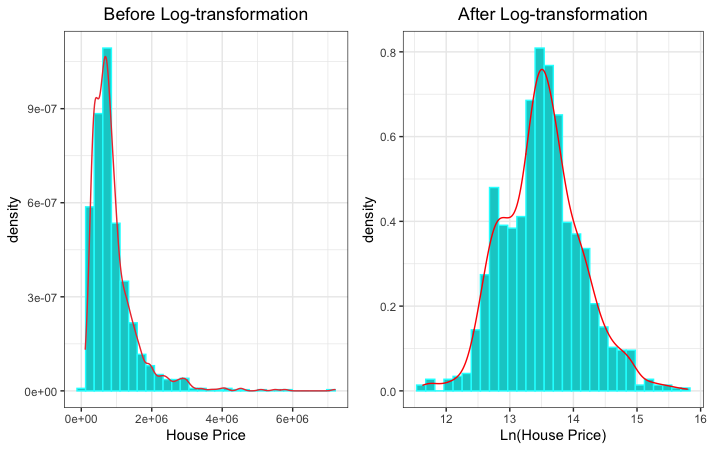}
\caption{Distribution of House Prices before and after Applying the Log-transformation.}\label{Figure1}
\end{figure}

The hedonic housing price regression model is essentially a function of a bundle of property characteristics such as structural attributes, socio-economic status of the neighbourhood, environmental amenities, and location. The hedonic housing price method is constructed by regressing the price of dwellings on a vector of housing attributes. Various functional forms have been applied in the literature to estimate the hedonic-price function including linear, semi-log, log-log, quadratic, and Box-Cox transformations. Here, we estimate hedonic price models for sold houses using a log-log specification. The advantage of a log-log specification over other models is that the log transformation of the dependent variable (house price) often mitigates heteroskedasticity. The log transformation reduces the problem of heteroskedasticity associated with the use of the highly-skewed sales price variable. Figure \ref{Figure1} plots the distribution of house prices before and after applying the log-transformation. Another important advantage of the log-log model is in obtaining a constant elasticity. In a log-log model the coefficients for those logarithmic terms are interpreted as elasticities. For example, we do log transformation on the regressor \qq{Living Area} to overcome non-linearity problems. The estimated coefficient on \qq{{Ln}(\text{Living Area})} measures the elasticity of property value with respect to the size of living area. In a log-log model a coefficient on a dummy variable such as a swimming pool that takes the value of $1$ if the house has one and zero otherwise, represents on average an approximate percentage difference in the prices of houses that have a swimming pool compared to those that do not. \\

The hedonic model using a log-log specification for houses sold over a year can be  written as follows:\\

$\text{Ln} (P)=  \mathbf{X} \beta+ \epsilon$ \\

\noindent where  $P$ is a [N $\times$ 1] vector and denotes house prices. $X$ is a [N × K] matrix of K independent variables describing the quantities of K characteristics of a property’s structure, neighbourhood characteristics, and location. $\epsilon$ is a [N $\times$ 1] vector of random error terms and $\beta$ is a corresponding [K $\times$ 1] vector of unknown parameters named regression coefficients. By using an ordinary least squares estimation method (OLS), the regression coefficients are estimated as:\\

$\hat{\beta} = (X^TX)^{-1}X^T Y$\\

The price of any residential property is affected by many property factors, namely, its structural, neighbourhood, and spatial attributes. Omitting important attributes from the hedonic regression model might bias estimates of implicit prices. To avoid misspecification biases a wide range of physical, neighborhood, locational and accessibility factors are controlled in the above hedonic model of residential property value. Here, to capture the structural attributes of the properties features such as number of bedrooms, bathroom, parking, and HOA fees (the annual fees for community costs) are included in our model. The dummy variables are used to represent the presence or absence of particular physical features and amenities such as a swimming pool, sauna, jacuzzi, or tennis court. The value 1 indicates the presence of a swimming pool, jacuzzi, or tennis court; a value of 0, represents otherwise. Other variables including house value, size of the living area, and lot area, are defined in natural logarithmic forms. The age of a property (in years) is computed from its date of completion to the year sold. The relationship between price and age is not expected to be linear. We are capturing the non-linear association between age and the house price by adding a quadratic term for property age. To explore the influence of propertie's distance to the important amenities namely employment centers, shopping centers, and schools we made use of Google Maps. For instance, to measure the accessibility of the property to employment centers and shopping centers we use driving time to the Boulder central business district (CBD) in downtown measured in minutes. In addition, to measure the accessibility of a house to the educational establishments we are adding the walking distance to primary, middle, and high school measured in time (minutes). To include the quality of the property's neighborhood we are adding school ratings; primary School, middle School, and high School Rates as well as the crime rate to our model. In addition, we are entering the socio-economic variables such as median household income and population density to our model.\\

\section{Data}

The data set we collected in this study to estimate the hedonic price model consists of $1061$ residential properties sold in the city of Boulder, Colorado. To isolate the influence of time on property prices the data used in this study is restricted to houses sold in a single year between January 1, 2019 and December 31, 2019 (see \cite{eckert1990property}). Samples used for hedonic estimation are not necessarily random draws from the population of dwellings, but are recently sold properties. \cite{robinson1979housing} states that \qq{recent sales are not necessarily a random draw from the total housing stock. If the purpose is to index the market of available units, this may not be of great concern, but if the purpose is to index the total stock, we must concern ourselves with possible selection bias}. Several studies have tested the existence of such biases. So far most studies have found the magnitude of the bias to be modest (see e.g., \cite{gatzlaff1998sample}).\\  

We collected real estate datasets from different resources consisting of Multiple Listing Service database\footnote{\url{https://realtyna.com/blog/list-mls-us}}, Public School Ratings\footnote{\url{https://www.greatschools.org}}, Colorado Crime Rates and Statistics Information\footnote{\url{https://www.neighborhoodscout.com/co/crime}}, CrimeReports\footnote{\url{https://www.cityprotect.com}}, WalkScore\footnote{\url{https://www.walkscore.com}}, and US Census Bureau\footnote{\url{https://data.census.gov/cedsci}}. Before proceeding with the estimation of our models we merged all data sets obtained from these websites. We screened our collected data set. Among these transactions 4 houses were in rough shape and needed to be rebuilt. Consequently, we deleted those 4 observations. All houses except one; which has lots of luxury furniture, have been sold unfurnished. So, we dropped the only furnished property. 30 properties were reported as horse properties. We eliminated those associated records. We checked for duplicate records. We found 4 duplicate transactions and eliminated those. In a number of observations we faced missing data points for some characteristics. For example, in some records the information for variables \qq{Bedroom}, \qq{Full Bathroom}, \qq{Parking}, \qq{Lot Area}, \qq{HOA Fees}, \qq{Solar Power} and \qq{Pool, Bath tub, Sauna, or Jacuzzi} were missing. We updated some of these missing data points by visiting different websites and looking up for the correct information. Finally, a few observations left with missing data points for the continuous variable \qq{Lot Area} and the dummy variables \qq{Solar Power} and \qq{Pool, Bath tub, Sauna, or Jacuzzi}. To deal with missing data points we could eliminate from the model any attributes that have missing values, or alternatively we could drop all those properties with missing attributes. However, deleting observations due to an incomplete list of attributes can cause sample size reduction and sample selection bias (see e.g., \cite{hill2013hedonic}). Instead, to avoid losing the samples from analysis we imputed missing values in the variable \qq{Lot Area} with its mean. Similarly, we imputed missing values for the dummy variables \qq{Solar Power} and \qq{Pool, Bath tub, Sauna, or Jacuzzi} with their mode. Another problem we encountered is the presence of outliers. The outliers were identified by calculating the $95\%$ percentile. To mitigate the influence of outliers on the analysis, we applied one-sided $95\%$ or two-sided $90\%$ Winsorization. That means outliers were set to the value of the $95\%$ percentile. The data cleaning process left us with a final sample of 1018 observations. The structural, spatial, and neighborhood attributes that are used in our analysis, are summarized in Table \ref{Table1} and \ref{Table2}. In addition to the mean, standard deviation, minimum and maximum, information about the relative standard deviation (the coefficient of variation ($CV=\frac{\sigma}{\mu}$), which represents the extent of variability in relation to the average of the variable has been reported in Table \ref{Table1}. \\

\begin{table}[h!]
\caption{Descriptive Statistics for Numerical Variables.}\label{Table1}
\centering
\begin{adjustbox}{max width=\textwidth}
\begin{tabular}{@{}l l l l l l@{}}\hline\hline\toprule\\
\multirow{2}[3]{*}{Variables} & \multicolumn{5}{c}{Aggregate Level}  \\\\
\cmidrule(lr){2-6}\\

&Mean & St. Dev.&Min &Max & Coefficient of Variation (CV) \\ [0.5ex] %
\hline
\hline
\\
$\text{House Price}$ ($\$$)& $896,332$ & 679,195.8 & 112,897 & 7,200,000 &$76\%$\\

$\text{Ln} (\text{House Price})$& 13.49 & 0.57 & 12.18 & 14.57&$4\%$\\

$\text{Ln} (\text{Lot Area})$ (SqFt) & 6.21 & 4.37 & 0.00 & 14.27 & 70\%\\

$(\text{Ln} (\text{Lot Area}))^2$ & 57.67 & 43.97 & 0.00 & 203.68& 76\% \\ 
Ln(Living Area) (SqFt)& 7.54 & 0.61 & 6.03 & 9.25  &8\%\\

$(\text{Ln} (\text{Living Area}))^2$ & 57.27 & 9.24 & 36.37 & 85.47&16\%\\

Age (year) & 43.12 & 21.46 & 1 & 98&50\%\\

$\text{Age}^2$  & 2,319.19 & 2,253.79 & 1 & 9,604 & 97\%\\


Full Bathroom  & 1.55 & 0.76 & 0 & 3& 49\% \\

Half Bathroom  & 0.41 & 0.51 & 0 & 2& 124\%\\ 

$\frac{3}{4}$ Bathroom  & 0.64 & 0.69 & 0 & 2& 108\%\\

Parking & 1.68 & 0.71 & 0 & 3& 42\%  \\ 

HOA Fees (annually) ($\$$)  & 1,693.32 & 2,033.16 & 0 & 7,113& 120\%\\ 

Drive to CBD (minute)  & 11.42 & 6.91 & 1 & 26& 61\% \\

Walk to E.School (minute) & 21.37 & 17.31 & 2 & 68& 81\% \\ 

Walk to M.school (minute) & 33.21 & 27.72 & 2 & 96  & 83\%\\ 

Walk to H.school (minute) & 46.94 & 32.92 & 4 & 122 & 70\%\\ 

Married ($\%$) & 42.97 & 16.87 & 9.90 & 70.30& 40\%\\

Median Household Inc. ($\$$)  & 61,137.44 & 20,891.56 & 19,985 & 96,406&  34\%\\

Neighborhood's Population & 43,641.85 & 46,872.42 & 888 & 99,081 & 107\%\\\\
\hline
Sample size  & 1018   \\ 
 [1ex]

\hline \hline
\end{tabular} 
\end{adjustbox}
\label{table:nonlin}
\end{table}

\begin{table}[!hp]
\caption{Descriptive Statistics for Categorical Variables.}\label{Table2}
\centering
\begin{adjustbox}{max width=\textwidth}
\begin{tabular}{@{}l l l l  l@{}}\hline\hline \toprule\\
\multirow{2}[3]{*}{Variables} & \multicolumn{4}{c}{Aggregate Level}  \\\\
\cmidrule(lr){2-5}\\

& Levels& Description & Frequency & Percent \\ [0.5ex] 
\hline
\hline\\
& 0 &No& 724 &71.12 \\[-1ex]  

\raisebox{1ex}{Pool, Bath tub, Sauna, or Jacuzzi} &  1&Yes&360&35.36 \\[1.5ex]

& 0&No & 658 &64.64\\[-1ex]  

\raisebox{1ex}{Solar Power} &  1&294&28.88&Yes\\[1.5ex]
 
& 1&A & 334 &32.81  \\[-1ex]  

\raisebox{1ex}{Nearest E.School Rank} & 2  &B
&548 &53.83  \\

& 3 &C &136 &13.36\\[1.5ex]
 
& 1 &A& 125&12.28 \\[-1ex]  

\raisebox{1ex}{Nearest M.School Rank } & 2  &B
&633 &62.18 \\

& 3& C &260 &25.54 \\[1.5ex]
 
\raisebox{1ex}{Nearest H.School Rank} & 1 &A
&1018 &100 \\[1.5ex]

& 1&Central & 230&22.59 \\[.01ex] 

& 2&North 
&238 &23.38  \\[-.05ex] 
& 3& South &143 &14.05 \\[-1ex] 
\raisebox{1ex}{Region} & 4& East&200 &19.65 \\

& 5& Gunbarrel&125 &12.28 \\

& 6& Rural&82 &8.06 \\[1.5ex]

& 0 &0 bedroom & 3&0.29\\[.01ex] 

& 1 &1 bedroom & 72&7.07\\[-.05ex] 

& 2&2 bedrooms &253 &24.85 \\ [-1ex] 

\raisebox{1ex}{Number of Bedrooms} 
& 3& 3 bedrooms&285 &28.00 \\

& 4& 4 bedrooms&249 &24.46 \\

& 5& 5 bedrooms&127 &12.48 \\

& 6& 6 bedrooms&27 &2.65 \\

& 7& 7 bedrooms&2 &0.20 \\[2.4ex]

& 1 &Condominum & 324&31.83 \\[-1ex]  

\raisebox{1ex}{Property Type} & 2&Town - Home 
&90 &8.84  \\

& 3& Single Family &604 &59.33 \\[1.5ex]

& 1 &Highest crime rate&82&8.06\\[-.75ex]  

\raisebox{1ex}{Neighborhood's Crime Level} & 2  &Middle crime rate&505&49.61\\

& 3& Lowest crime rate & 431 &42.34\\[1.75ex]
 
 \hline
Sample size&-&1018&100&-\\

\hline \hline
\end{tabular} 
\end{adjustbox}
\label{table:nonlin}
\end{table}

\begin{figure}[h!]
\centering
\hspace*{.00001\textwidth} 
\includegraphics[width=0.95\textwidth ,height=0.45\textheight]{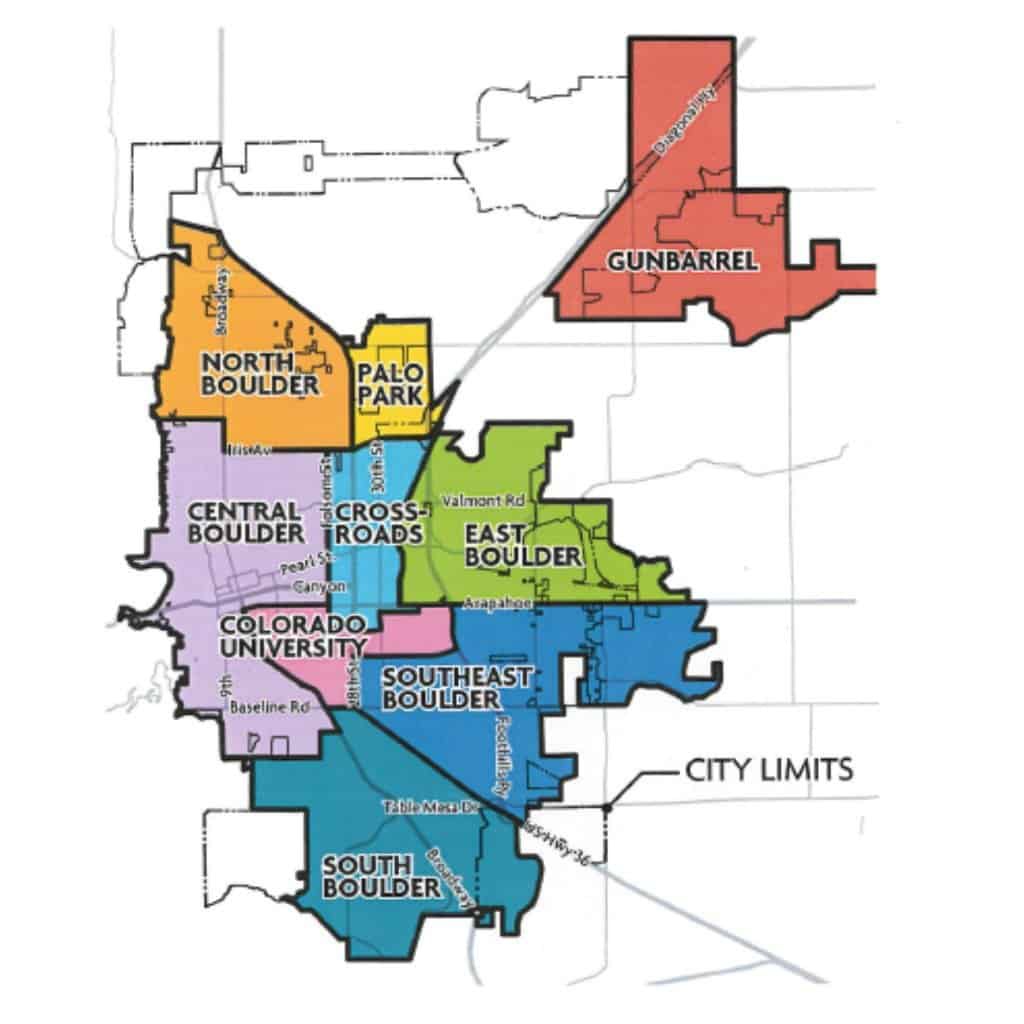}
\caption{City of Boulder.}\label{Figure2}
\end{figure}

\begin{figure}[t]
 \makebox[\textwidth][c]{\includegraphics[width=19cm,height=0.52\textheight]{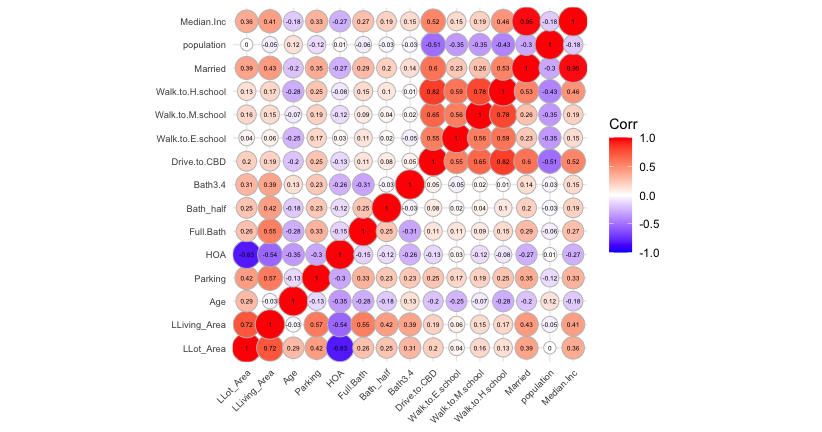}}
\caption{Correlation Plot.}\label{Figure3}
\end{figure}

City of Boulder is divided into seven different geographical locations; central Boulder, downtown Boulder, old north Boulder, north Boulder, south Boulder, east Boulder, Gunbarel, and rural areas. Figure \ref{Figure2} plots the city of Boulder on the map. With city development, the old north Boulder neighborhood is in central Boulder. Therefore, we are merging the data for the central Boulder and old north Boulder. We explored the geographical location of each property by making use of Google Maps. For example, in the north Boulder region there are $238$ transactions. $200$ dwellings sold in the east Boulder. There are only $49$ transactions in downtown Boulder. Because the number of transactions in downtown Boulder is limited we are merging the data for the central Boulder and downtown Boulder. Finally, we are considering six different geographical areas in the city of Boulder as central Boulder, north Boulder, south Boulder, east Boulder, Gunbarel, and rural areas. Table \ref{Table5} in the appendix provides more information about the number of transactions in different geographical locations in $2019$. \\

The property types in the housing market in the city of Boulder are classified as condominiums, town-homes, and single-family houses. The town-home properties in the city of Boulder have been classified into the west of the Broadway highway and downtown, and the east of the Broadway highway and all other areas in the city of Boulder\footnote{\url{https://www.bouldercounty.org/property-and-land/assessor/sales/comps-2019/residential}}. There are only $22$ town-homes to the west of the Broadway highway and downtown and $69$ town-homes to the east of the Broadway highway and all other areas in the city of Boulder. Small number of observations makes the creation of reliable and robust models difficult. \cite{malpezzi2002hedonic} states that if sample sizes are small, it is normally best to avoid disaggregating the market. Consequently, we are combining the sold town-homes to the west and east of the Broadway highway and the downtown.\\

Before proceeding with our models we should test for multicollinearity problems, the common issue in hedonic regression models. In the presence of high correlation the estimated coefficients may be unreliable (see e.g., \cite{sheppard1999hedonic}). Due to high correlation, the corresponding coefficients might be estimated as statistically insignificant when in fact they are statistically significant. Consequently this problem poses an inaccurate interpretation of coefficients and makes it difficult to recognize important regressors. To ensure the robustness of our model, the Pearson correlation coefficient is computed. Figure \ref{Figure3} summarizes this information. Positive correlations are shown in red color and negative correlations in dark blue color. The color intensity and the size of the circle are proportional to the correlation coefficients. We drop variables which generate pairwise correlation coefficients greater (smaller) than the common thresholds $0.8$ (-$0.8$). In the citywide housing market we observed that variables \qq{Walk to H.school}, \qq{Drive to CBD}, \qq{Married} and \qq{Median Household Inc.} are highly correlated. To avoid multicollinearity problems we dropped \qq{Walk to H.school} and \qq{Married} and calculated the Pearson correlation coefficient one more time. In addition, we calculated the generalized variance inflation factors (GVIF). The accepted GVIF cutoff is $\sqrt{VIF}=\sqrt{5}=2.24$. In the aggregate level data set no GVIF is above the defined thresholds. Therefore, we keep the rest of the variables in the model.\\

\section{Aggregate and Disaggregate Market Levels}

We first estimate the hedonic price equation at the citywide market level; over all 1018 houses sold in the city of Boulder in 2019. To control for property type and location differences we are adding categorical variables into the hedonic pricing model. We use a non-stepwise regression method and keep both statistically significant and non-statistically significant regressors in our regression model. An underlying problem with stepwise regression is that in the presence of heteroscedasticity, the results of the t-tests are unreliable. Consequently, some regressors that have causal effects on the dependent variable may not be statistically significant, although some other regressors may be significant by chance. Therefore, using a stepwise regression method may fit the model well over our sample (overfitting), but nonetheless, do poorly over an unseen data set (see e.g., \cite{smith2018step}). Initially, we are employing an ordinary least square (OLS) estimation method to provide information of the attributes affecting the housing market in Boulder, Colorado. Table \ref{Table3} summarizes this information.\\

\begin{table}[!hp]
\caption{Hedonic Model Estimates\\OLS Method.\\}\label{Table3}
\centering 
\ra{1.3}
\begin{adjustbox}{max width=\textwidth}
\begin{tabular}{@{}lllllrrr@{}}\toprule
\hline
\multirow{2}[3]{*}{Independent Variables} & \multicolumn{3}{c}{Aggregate Level} \\ \cmidrule{2-4} 
& Coeff. & Std. Error& T-Statistics\\ \midrule
\hline\\
(Intercept) & $13.49$ & $1.65$ & $8.20^{***}$ \\ 

Ln(Lot Area) & $0.38$ & $0.06$ & $5.79^{***}$ \\

$(\text{Ln} (\text{Lot Area}))^2$ & -$0.004$ & $0.003$ & -$1.36$ \\ 

Ln(Living Area) & -$0.48$ & $0.45$ & -$1.08$  \\

$(\text{Ln} (\text{Living Area}))^2$ & $0.07$ & $0.03$ & $2.30^{**}$ \\

Age & -$0.02$ & $0.01$ & -$3.95^{***}$ \\

$\text{Age}^2$ & $0.0001$ & $0.0000$ & $7.39^{***}$  \\

\textbf{Number of Bedrooms}\\

(Bedrooms)2 & $0.06$ & $0.04$ & $1.73^{*}$ \\

(Bedrooms)3 & $0.02$ & $0.05$ & $0.52$ \\ 

(Bedrooms)4 & -$0.02$ & $0.05$ & -$0.46$ \\ 

(Bedrooms)5+ & -$0.05$ & $0.06$ & -$0.96$ \\ 

Full Bathroom & $0.08$ & $0.02$ & $4.69^{***}$ \\

Half Bathroom & $0.01$ & $0.02$ & $0.85$ \\ 

$\frac{3}{4}$ Bathroom & $0.06$ & $0.02$ & $3.54^{***}$\\

Parking & $0.04$ & $0.01$ & $3.12^{**}$  \\ 

HOA Fees & $0.0000$ & $0.0000$ & $1.74^{*}$\\

\textbf{School Ranking}\\

(Nearest E.School Rank)2 & -$0.06$ & $0.02$ & -$2.84^{***}$ \\

(Nearest E.School Rank)3 & -$0.17$ & $0.03$ & -$5.04^{***}$ \\ 

(Nearest M.School Rank)2 & -$0.18$ & $0.04$ & -$4.06^{***}$ \\ 

(Nearest M.School Rank)3 & -$0.14$ & $0.05$ & -$2.76^{***}$ \\ 

(Pool, Bath tub, Sauna, or Jacuzzi)1 & $0.04$ & $0.02$ & $2.21^{**}$ \\

(Solar Power)1 & $0.09$ & $0.03$ & $3.17^{***}$ \\

Drive to CBD & -$0.02$ & $0.003$ & -$7.32^{***}$ \\

Walk to E.School & -$0.0003$ & $0.001$ & -$0.48$ \\ 

Walk to M.school & -$0.001$ & $0.001$ & -$1.45$\\ 

Median Household Inc. & -$0.0000$ & $0.0000$ & -$0.62$\\

Neighborhood's Population& -$0.0000$ & $0.0000$ & -$1.21$\\
\hline \hline
\end{tabular}
\end{adjustbox}
\end{table}

\begin{table*}
\ContinuedFloat
\caption{(Continued.)}
\ra{1.3}
\begin{adjustbox}{max width=\textwidth}
\begin{tabular}{@{}llllllrr@{}}\toprule
\hline
\multirow{2}[3]{*}{Independent Variables} \thickspace \thickspace& \multicolumn{3}{c}{Aggregate Level} \\ \cmidrule{2-4} 
& \thickspace Coeff. \thickspace & Std. Error\thickspace& T-Statistics \\ \midrule
\hline

\textbf{Crime Levels} (Base: Level 3) \\

Neighborhood's Crime Level: 1 \thickspace \thickspace \thickspace \thickspace \thickspace \thickspace \thickspace \thickspace \thickspace& -$0.09$ & $0.03$ & -$2.61^{***}$\\ 

Neighborhood's Crime Level: 2  \thickspace & -$0.03$ & $0.02$ & -$1.70^{*}$ \\ 

\textbf{Property Type}\\

Single Family Houses & -$1.05$ & $0.24$ & -$4.35^{***}$ \\

Town-Home Houses & -$1.10$ & $0.23$ & -$4.84^{***}$\\

\textbf{Regions} (Base: Central)\\

(North)2 & -$0.16$ & $0.03$ & -$4.88^{***}$  \\ 

(South)3 & -$0.14$ & $0.05$ & -$3.00^{***}$ \\ 

(East)4 & -$0.22$ & $0.04$ & -$6.26^{***}$\\ 

(Gunbarrel)5 & -$0.18$ & $0.06$ & -$3.01^{***}$  \\ 

(Rural)6 & -$0.30$ & $0.06$ & -$4.94^{***}$\\ 

Ln(Lot Area):Ln(Living Area) & -$0.02$ & $0.01$ & -$3.70^{***}$\\

Ln(Living Area):Age  & $0.001$ & $0.001$ & $2.01^{**}$  \\\\
\hline 

Sample size& 1018 \\ 

$\text{R}^2$ &  $0.865$\\

$\text{Adj. R}^2$ & $0.86$\\

F-statistic ($p-$value)& $169.7$&($0.000$)\\
Dependent variable: Ln(Price)\\\bottomrule
\hline \hline
\end{tabular}
\end{adjustbox}
\caption*{\textbf{Note}: $***,~ **$, and $*$ ~indicate $p < 0.01$, $p < 0.05$, and $p < 0.1$ respectively.\\ Base group in the citywide housing market: condominium houses, with one bedroom or less, no pool, bath tub, sauna, or jacuzzi, no solar power, in the central Boulder region with crime level 3. }
\end{table*}

The urban property markets contain high variation in the structural, locational, neighbourhood, and environmental attributes. We both graphically and statistically detected heteroskedasticity in the aggregate market level. In the presence of heteroskedasticity OLS estimators are unbiased, while the standard error and the statistics we used to test hypotheses based on ordinary least square estimation method will be inefficient. Following \cite{fletcher2000heteroscedasticity} we use White’s standard errors test to correct the issue of incorrect standard errors and test statistics. The robust standard errors and heteroskedasticity-consistent estimates are reported in Table \ref{Table4}.\\ 

In the presence of heteroscedasticity a single model over the pooled property data set is not reliable. A model over the aggregate housing data set is assessing the whole city and would hold in high variation in houses characteristics. Less difference in house attributes lead to a more homogeneous market which may provide more efficient and reliable estimates. In the light of these facts, we stratify the market into a number of distinct submarkets to create a number of homogeneous submarkets from the aggregate heterogeneous dwelling data set. Normally, a submarket composed of groups of relatively homogenous dwellings that are recognised as close substitutes to potential house buyers (see e.g., \cite{goodman1998housing}, \cite{bourassa2003housing}, \cite{taylor2008theoretical}, \cite{paez2008moving}, \cite{wheeler2014ab}, and \cite{wu2020analyzing}). There is no consistent definition for housing submarkets. Some researchers suggested that housing submarkets compose of all houses with similar environmental characteristics. Others have defined dwelling submarkets based on socio-economic attributes (see e.g., \cite{palm1978spatial}). Some analysts have defined dwelling submarkets based on the structural characteristics (see e.g., \cite{watkins1999property}, \cite{adair2000house}, and \cite{carriazo2018demand}). Others have defined dwelling submarkets as a collection of houses within a specific location (see e.g., \cite{straszheim1975front}, and \cite{keskin2017defining}). Interestingly, some other researchers have realized the importance of both structural and locational factors in determining housing market segmentations (see e.g., \cite{adair1996hedonic}, \cite{watkins2001definition}, \cite{alkana2015housing}, \cite{xiao2016urban}, and \cite{bello2020conventional}).\\

\begin{table}[!h]
\caption{Hedonic Model Estimates\\White Correction.\\}\label{Table4}
\centering 
\ra{1.3}
\begin{adjustbox}{max width=\textwidth}
\begin{tabular}{@{}lllllrr@{}}\toprule
\hline
\multirow{2}[3]{*}{Independent Variables} & \multicolumn{3}{c}{Aggregate Level} \\ \cmidrule{2-4} 
& Coeff. & White’s Std. Error& White’s T-Statistics\\ \midrule
\hline\\
(Intercept) & $13.49$ & $1.62$ & $8.32^{***}$ \\ 

Ln(Lot Area) & $0.38$ & $0.10$ & $3.78^{***}$ \\

$(\text{Ln} (\text{Lot Area}))^2$ & -$0.004$ & $0.005$ & -$0.85$ \\ 

Ln(Living Area) & -$0.48$ & $0.44$ & -$1.10$  \\

$(\text{Ln} (\text{Living Area}))^2$ & $0.07$ & $0.03$ & $2.34^{**}$ \\

Age & -$0.02$ & $0.01$ & -$3.00^{***}$  \\

$\text{Age}^2$ & $0.0001$ & $0.0000$ & $5.73^{***}$ \\

\textbf{Number of Bedrooms}\\

(Bedrooms)2 & $0.06$ & $0.04$ & $1.71^{*}$ \\

(Bedrooms)3 & $0.02$ & $0.05$ & $0.49$ \\ 

(Bedrooms)4 & -$0.02$ & $0.06$ & -$0.44$ \\ 

(Bedrooms)5+ & -$0.05$ & $0.06$ & -$0.89$\\ 

Full Bathroom & $0.08$ & $0.02$ & $4.68^{***}$ \\

Half Bathroom & $0.01$ & $0.02$ & $0.80$ \\ 

$\frac{3}{4}$ Bathroom & $0.06$ & $0.01$ & $3.76^{***}$\\

Parking & $0.04$ & $0.01$ & $2.74^{***}$  \\ 

HOA Fees & $0.0000$ & $0.0000$ & $1.38$\\

\textbf{School Ranking}\\

(Nearest E.School Rank)2 & -$0.06$ & $0.02$ & -$2.67^{***}$ \\

(Nearest E.School Rank)3 & -$0.17$ & $0.03$ & -$5.27^{***}$ \\ 

(Nearest M.School Rank)2 & -$0.18$ & $0.04$ & -$4.35^{***}$\\ 

(Nearest M.School Rank)3 & -$0.14$ & $0.05$ & -$2.83^{***}$ \\ 

(Pool, Bath tub, Sauna, or Jacuzzi)1 & $0.04$ & $0.02$ & $2.08^{**}$ \\

(Solar Power)1 & $0.09$ & $0.03$ & $2.46^{**}$ \\

Drive to CBD & -$0.02$ & $0.003$ & -$5.94^{***}$ \\

Walk to E.School & -$0.0003$ & $0.001$ & -$0.47$ \\ 

Walk to M.school & -$0.001$ & $0.001$ & -$1.22$\\ 

Median Household Inc. & -$0.0000$ & $0.0000$ & -$0.53$\\

Neighborhood's Population& -$0.0000$ & $0.0000$ & -$1.08$\\
\hline \hline
\end{tabular}
\end{adjustbox}
\end{table}

\begin{table*}[t]
\ContinuedFloat
\caption{(Continued.)}
\ra{1.3}
\begin{adjustbox}{max width=\textwidth}
\begin{tabular}{@{}llllcrrr@{}}\toprule
\hline
\multirow{2}[3]{*}{Independent Variables} & \multicolumn{3}{c}{Aggregate Level} \\ \cmidrule{2-4} 
& Coeff. & White’s Std. Error& White’s T-Statistics\\ \midrule
\hline

\textbf{Crime Levels} (Base: Level 3) \thickspace  \thickspace\\

Neighborhood's Crime Level: 1 & -$0.09$ & $0.03$ & -$2.65^{***}$\\ 

Neighborhood's Crime Level: 2 & -$0.03$ & $0.02$ & -$1.61$ \\ 

\textbf{Property Type}\\

Single Family Houses & -$1.10$ & $0.41$ &-$2.68^{***}$ \\

Town-Home Houses & -$1.05$ & $0.43$ & -$2.45^{**}$\\

\textbf{Regions} (Base: Central)\\

(North)2 & -$0.16$ & $0.04$ & -$4.36^{***}$   \\ 

(South)3 & -$0.14$ & $0.05$ & -$3.02^{***}$\\ 

(East)4 & -$0.22$ & $0.04$ & -$5.48^{***}$\\ 

(Gunbarrel)5 & -$0.18$ & $0.07$ & -$2.55^{**}$ \\ 

(Rural)6 & -$0.30$ & $0.07$ & -$4.33^{***}$\\ 

Ln(Lot Area):Ln(Living Area) & -$0.02$ & $0.01$ & -$3.37^{***}$\\

Ln(Living Area):Age  & $0.001$ & $0.001$ & $1.65^{*}$ \\ \\

\hline

Chi-squared ($p-$value)& $115$&($0.002$)\\
Dependent variable: Ln(Price) \\\bottomrule
\hline \hline
\end{tabular}
\end{adjustbox}
\caption*{\textbf{Note}: $***,~ **$, and $*$ ~indicate $p < 0.01$, $p < 0.05$, and $p < 0.1$ respectively.\\}
\end{table*}

To test for market segmentation we first stratify the housing market in the city of Boulder based on both property type and location. Next, we nest housing submarkets based on property type within spatially defined housing market segmentation. \\

To avoid violating the classical linear model assumptions we estimate the hedonic price regression model for each submarket separately using the OLS estimation methods. In the submarkets based on property type to control for location differences categorical variables are added into the hedonic pricing models. Similarly, in the submarkets based on spatial attributes to control for property type differences categorical variables are included into the hedonic pricing models. In a number of submarkets due to the limited number of houses, some of the categorical variables were removed.\\

Next, we perform Chow test to examine if there is significant difference between hedonic price regression models across distinct submarkets (see e.g., \cite{watkins1999property}, \cite{yu2006studies}, \cite{taylor2008theoretical}, \cite{gavu2019empirical}, and \cite{yang2020place}). \\

Finally, we nest structural submarkets within spatial submarkets and focus on the joint importance of both structural and locational attributes in classifying submarkets. \\

\section{Results}

In this section we first provide summary statistics about the houses in the aggregate level. Among the 1018 residential properties sold during 2019 in the city of Boulder, the asset prices range between $\$112,897$ and $\$7,200,000$, with an average price of $\$896,332$ and a median price of $\$729,500$. The average \qq{Living Area} and \qq{Lot Area} are $2,264$ and $18,367$ square feet respectively. The average age of the dwellings is approximately $43$ years old. In this data set the houses have $3.16$ bedrooms on average with $0.29\%$ of them having no bedroom and $15.32\%$ of them having 5 bedrooms or more. Moreover, $28.88\%$ of the dwellings have solar power and $35.36\%$ of them have a pool, bath tub, sauna, or jacuzzi. More information about the descriptive statistics of the variables in the citywide housing market level can be found in Table \ref{Table1} and \ref{Table2}.\\

Next, we bring descriptive statistics about the houses in the disaggregate level. Amongst the $1018$ dwellings in our data set, $604$ are single family houses, $324$ are condominiums, and $90$ are town-homes. The single family property prices range between $\$216,575$ and $\$7,200,000$, with an average price of $\$1,160,321$. In the town-home submarket the house prices range between $\$115,000$ and $\$1,421,000$, with an average price of $\$627,960$. In the condominium submarket the dwelling prices range between $\$112,897$ and $\$2,600,000 $, with an average price of $\$478,751$. In the single family housing submarket the dwellings have $3.82$ bedrooms on average with $0.17\%$ of them having no bedroom and $24.51\%$ of them having $5$ bedrooms or more. The houses in the town-home submarket have $2.97$ bedrooms on average with $7.78\%$ of them having $5$ bedrooms or more. In the condominium submarket the houses have $1.98$ bedrooms on average with $0.62\%$ of them having no bedroom and $0.31\%$ of them having $5$ bedrooms or more. In the single family housing submarket $3.31\%$ of the houses have solar power and $18.05\%$ of the houses have a pool, bath tub, sauna, or jacuzzi. We do not have much information about the existence of solar power, pool, bath tub, sauna, or jacuzzi in the town-home or condominium submarkets.\\

In the locational submarkets the average age of the dwellings are almost $30$ years old in north Boulder, $59$ years old in central Boulder, $53$ years old in south Boulder, $40$ years old in east Boulder, $37$ years old in Gunbarrel, and $41$ years old in the rural region. In addition, in the spatial submarkets, namely, north Boulder region with $238$ transactions the house prices range between $\$134,306$ and $\$4,500,000$, with an average price of $\$807,214$. Amongst the $230$ dwellings sold in the central Boulder the house values range between $\$115,000$ and $\$7,200,000$, with an average price of $\$1,256,235$. Table \ref{Table5} in the appendix provides more information about the descriptive statistics of the house prices in different submarkets. From Table \ref{Table5} we learn that the deviation in residential property prices is lower in the north, south, east, Gunbarrel submarkets, than the aggregate market level. However, the house price difference is higher in central Boulder and rural areas. \\

Given the descriptive statistics of the houses attributes we will now focus on the outcomes from performing the hedonic models in different submarket levels. Initially, to ensure the robustness of our models at the disaggregate levels, the Pearson correlation coefficients have been computed in each submarket. Namely, in the central Boulder submarket the variables \qq{HOA Fees} and \qq{Ln(Living Area)} are greatly correlated ($0.88$). In south Boulder, VIF for the regressors \qq{Median Household Inc.} and \qq{HOA Fees} is $16.27$ and $8.20$ respectively. Variables \qq{Ln(Lot Area)} and \qq{Ln(Living Area)} are highly correlated in Gunbarrel ($0.83$). In addition, the correlation between variables \qq{Drive to CBD} and \qq{Walk to E.School} is $0.81$. In the rural submarket the correlation between \qq{Walk to E.School} and \qq{Walk to M.School} is $0.91$. In consideration of these high correlations, to prevent multicollinearity problems we excluded these explanatory variables from the hedonic models in these submarkets.\\

Based on the outcomes from the hedonic models in different submarkets, we observe that the estimated marginal contribution of the residential property features on house values vary. However, in general, the sign of the marginal effect of the dwellings' characteristics are consistent with our expectations. For instance, as anticipated the higher the size of the lot area the higher the house prices are or the existence of solar power significantly raises the house value. Age of the property is negatively associated with the dwelling value, but the age-squared term is positively related to the house price. Moreover, commuting time to the CBD negatively impacts the property prices. Allocation of the dwellings in the neighborhoods with lower crime rate add to their values except in Gunbarrel and south Boulder. Additionally, the presence of a swimming pool, sauna, or jacuzzi add to the dwelling values. For more details see Table \ref{Table6} and \ref{Table7} in the appendix.\\ 

\begin{figure}[h]
 \makebox[\textwidth][c]{\includegraphics[width=10cm]{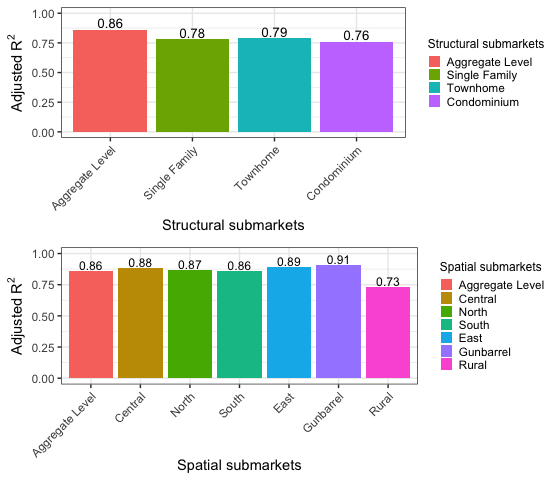}}
\caption{Comparison of the Performance of Hedonic Regression Models in Different Submarkets.}\label{Figure4}
\end{figure}

To compare the explanatory power of the hedonic regression models at the aggregate and disaggregate levels we rely on values of adjusted $R^2$ which are ranged from $0.73$ to $0.91$. The values of adjusted $R^2$ from the models in the structural submarkets are smaller than the model in the citywide level. This could be due to smaller sample sizes and fewer regressors in the models. Nevertheless, in the spatial submarkets the models in all submarkets, but rural submarket, have higher adjusted $R^2$ and higher explanatory powers than the model in the citywide level. The higher levels of explanation in the spatial submarkets might represent the importance of locational characteristics in determining housing market segmentations in the city of Boulder. More information about the explanatory power of these models can be learned from the bar-plots in Figure \ref{Figure4}. \\ 

\begin{figure}[t]
 \makebox[\textwidth][c]{\includegraphics[width=10cm]{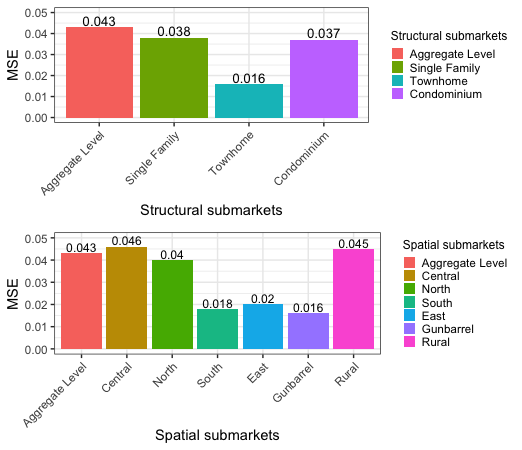}}
\caption{Comparison of the Mean Squared Errors in Different Submarket Levels with the Citywide Market Level.}\label{Figure7}
\end{figure}

In addition to measuring the explanatory power of the hedonic regression models, we checked for the homogeneity assumption in different models. We observed a higher level of homogeneity in most housing submarkets compared to the citywide housing market level. The $\emph{p}$-value from the Breusch-Pagan test is represented in Table \ref{Table8}. Moreover, to evaluate the performance of hedonic models over different market levels, we computed the performance metric, the mean squared error (MSE). The values of the MSE are lower for all submarkets except central Boulder and rural submarkets. Given the level of homogeneity in the housing submarkets and also the house price variability these outcomes have been expected. The values of MSE in the aggregate and disaggregate levels, are depicted in Figure \ref{Figure7}.\\

\begin{sidewaysfigure}[hp]
 \makebox[\textwidth][c]{\includegraphics[width=20cm]{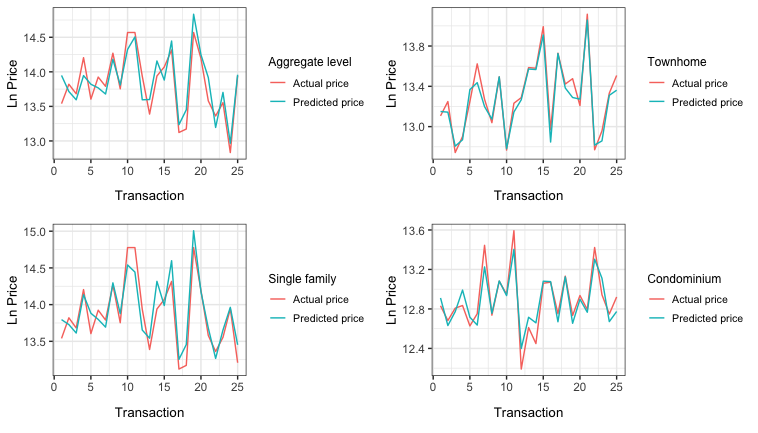}}
\caption{The Predicted Prices Obtained by Hedonic Regression Model in Structural Submarket and the Actual Prices.}\label{Figure5}
\end{sidewaysfigure}

Furthermore, to visualize the accuracy of our housing price prediction models we plotted dwellings' actual and predicted prices in the aggregate and disaggregate levels (Figure \ref{Figure5} and \ref{Figure6}). \\

\begin{sidewaysfigure}
 \makebox[\textwidth][c]{\includegraphics[width=20cm]{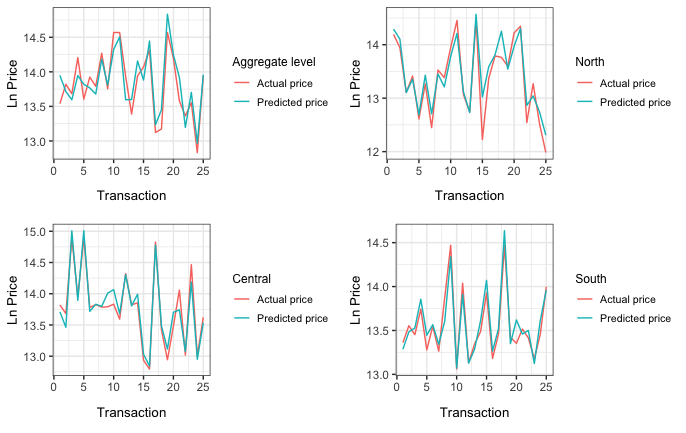}}
\caption{The Predicted Prices Obtained by Hedonic Regression Model in Spatial Submarket and the Actual Prices.}\label{Figure6}
\end{sidewaysfigure}

\begin{sidewaysfigure}
\ContinuedFloat
 \makebox[\textwidth][c]{\includegraphics[width=20cm]{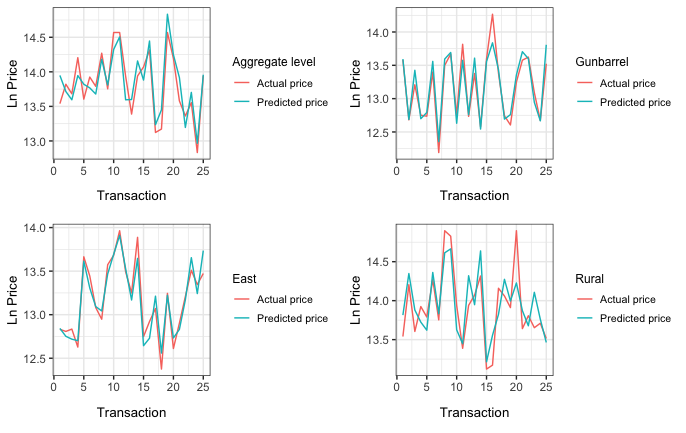}}
\caption{(Continued) - The Predicted Prices Obtained by Hedonic Regression Model in Spatial Submarket and the Actual Prices.}\label{Figure6}
\end{sidewaysfigure}

To examine if there are significant differences between hedonic models across distinct segments we performed Chow test. The results from Chow test are presented in table \ref{Table9} in the appendix. These results indicate that significant structural differences exist between single family and condominium submarkets. Similarly, we observe significant structural differences across single family and town-home segments. Nonetheless, the results from the chow test implies no sufficient evidence of market segmentation between town-home and condominium submarkets. Considering this, we merged these two submarkets and considered them as a single homogeneous market (condo-town-home).\\

Furthermore, the results from chow test, shown in table \ref{Table10} in the appendix confirms that spatial differences exist across most of the submarkets. For instance, the central Boulder differs from all other submarkets at $1\%$ significance level. The north and south Boulder submarkets vary from east Boulder. The east Boulder submarket differs from Gunbarrel. For those special submarkets that indicate no significant house price differences such as north and the rural area, south and Gunbarrel we combined these submarkets and named this merged submarket, north south Gunbarrel rural (NSGR) submarket. \\

Finally, we nested structural housing submarkets within spatial housing segmentations. We observed that in the nested single family home submarket within the NSGR submarket (NSGR-single family) the variables \qq{Walk to M.school} and \qq{Drive to CBD} are highly correlated $(0.83)$. The variable \qq{HOA Fees} is also highly correlated with \qq{Ln(Lot Area)} $(0.81)$. In the nested central Boulder submarket within the condo-town-home submarket (central-condo-town-home), GVIF for the regressor \qq{Neighborhood's Crime Level} is $2.49$. In the nested east and condo-town-home submarket variables \qq{Walk to M.school} and \qq{Drive to CBD} are highly correlated (-$0.94$). In addition, the GVIF for the categorical variable \qq{Nearest E.School Rank} is $3.83$. In the nested NSGR and condo-town-home submarket GVIF for variable \qq{Nearest E.School Rank} is $3.13$. In the nested NSGR and single family housing submarket GVIF for the categorical variable \qq{Nearest M.School Rank} is $3.26$. Consequently, to prevent multicollinearity problems we excluded these explanatory variables from the hedonic models in these nested submarkets. Table \ref{Table12} in the appendix represents a summary of results for the hedonic models in different nested submarkets. Furthermore, we compared hedonic price regression models for different nested submarkets using Chow test. Table \ref{Table13} in the appendix provides us with a summary of results.\\

\section{Conclusion}

In this paper we implemented hedonic price regression models, to predict house prices in the city of Boulder, Colorado. In the urban property markets the houses have heterogeneous characteristics. They differ in structural, locational, neighbourhood, and environmental attributes. Consequently, in urban housing markets stratification could improve the performance of housing price prediction models. In this study, we stratified the real estate market in the city of Boulder into the structural and spatial submarkets.\\ 

We compared several performance measurements for the predictions of the house prices in the aggregate and disaggregate market levels. We observe higher levels of homogeneity and explanatory power and lower values of the MSE in most housing submarkets compared to the citywide housing market. In addition, the results obtained from the Chow tests confirm the structural and spatial differences across most of the submarkets. Considering these performance metrics we are anticipating that the housing market in the city of Boulder should be necessarily classified into the structural and spatial submarkets.\\

The hedonic models have great advantages. One major supremacy of the hedonic model is its simplicity in interpreting the estimated parameters. Another advantage of the hedonic methods is that the marginal values of the attributes can be obtained by differentiating the price function with respect to each characteristic (see e.g., \cite{mcmillan1980extension}). Despite these advantages, hedonic models have been mainly applied on the residential real estate market and barely been implemented in other real estate markets. It would be interesting to perform hedonic models over non-residential real estate markets, including business, industrial, or commercial properties such as hotels, clubs, educational properties, hospitals, and farms.\\

In addition, it would be interesting to compare the performance of the hedonic approach with the machine learning and deep learning algorithms such as artificial neural network, random forest, k nearest neighbor, and support vector regression which have been offered for the house price valuation and real estate property price prediction model.\\

\section*{Acknowledgements}
Thanks goes to Dr. Nicholas E. Flores\footnote{Department Chair of Economics, Colorado University at Boulder} for his comments.

\newpage
\renewcommand\bibname{References}

\bibliographystyle{apalike}

\bibliography{mybib.bib}
\newpage

\pagenumbering{arabic}
\renewcommand*{\thepage}{A\arabic{page}}

\renewcommand\thefigure{\thesection.\arabic{figure}} 
\renewcommand\thetable{\thesection.\arabic{table}} 

\label{app:Online}

\begin{appendix}
\section{Appendix}

\begin{table}[!hp]
\caption{Summary Statistics of House Prices in Different Submarkets.\\}\label{Table5}
\captionsetup{singlelinecheck = false}
\centering 
\ra{1.3}
\begin{adjustbox}{max width=\textwidth}
\begin{tabular}{@{}llllllllllllll@{}}\toprule
\hline
\multirow{2}[3]{*}{Market Level} & \multicolumn{13}{c}{House Price} \\ \cmidrule{6-14} 
&&&&&& Mean && St. Dev.&&   Min && Max \\ \midrule
\hline\\
Citywide  &&&&&& $896,332$  && $679,195.8$  && $112,897$ && $7,200,000$\\
\\
Single Family &&&&&&$1,160,321 $&&$739,401.9$ &&$216,575$&&$7,200,000$\\
\\
Town-Home   &&&&&&$627,960$ &&$246,470$ &&$115,000$  && $1,421,000$\\
\\
Condominum &&&&&& $478,751$ &&$299,644.7$&&$112,897$ &&$2,600,000 $\\[-3ex]
{\color{white}your comment ...} &&&&&&&&&&&&&\\
Central &&&&&& $1,256,235$ &&$921,806.1$ &&$115,000$ && $7,200,000$ \\
\\
North &&&&&&$807,214$ &&$517,334$&&$134,306$&&$4,500,000$\\
\\
South&&&&&&  $920,577$ &&$528,865.8$&&$243,000$ &&$4,550,000$ \\
\\
East&&&&&&   $646,616$  &&$373,423.8$ &&$112,897$ &&$3,350,000$ \\
\\
Gunbarrel&&&&&& $593,812$&& $307,783.2$&&$194,585$&&$1,995,051$ \\
\\
Rural&&&&&&1,$173,444$ &&$929,252.6$&& $425,000$&&$5,779,000$\\\\[1ex]

\hline \hline
\end{tabular}
\end{adjustbox}
\label{table:nonlin}
\caption*{\textbf{Note}:  The house prices have been recorded in the US dollars ($\$$). }
\end{table}

\begin{sidewaystable} 
\setlength\tabcolsep{0pt}
\caption{Hedonic Model Estimates - Property Type Submarkets\\OLS Method.}\label{Table6}
\centering 
\ra{1.3}

\begin{adjustbox}{max width=\textwidth}
\begin{tabular}{@{}llllllllllll@{}}\toprule
\hline
\multirow{2}[3]{*}{Independent Variables} & \multicolumn{3}{c}{Single Family Housing} & \phantom{abc}& \multicolumn{3}{c}{Town-Home Housing} &
\phantom{abc} & \multicolumn{3}{c}{Condominium Housing}\\ \cmidrule{2-4} \cmidrule{6-8} \cmidrule{10-12}
& Coeff.  & Std. Error\thickspace \thickspace& T-Statistics&& Coeff.& Std. Error \thickspace \thickspace& T-Statistics && Coeff.& Std. Error \thickspace \thickspace& T-Statistics\\ \midrule
\hline\\

(Intercept) & $24.93$ & $2.93$ & $8.52^{***}$&& $10.73$ & $9.71$ & $1.11$  && $9.56$ & $4.71$ & $2.03^{**}$ \\ 

Ln(Lot Area) & $0.13$ & $0.28$ & $0.46$&&- & -& -&&-&-&- \\

$(\text{Ln} (\text{Lot Area}))^2$ & -$0.02$ & $0.01$ & -$1.85^{*}$&& -& -& -&&-&-&- \\ 

Ln(Living Area) & -$3.34$ & $0.65$ & -$5.10^{***}$ && $0.86$ & $2.64$ & $0.33$ &&$0.40$ & $1.30$ & $0.30$ \\

$(\text{Ln} (\text{Living Area}))^2$ & $0.19$ & $0.04$ & $4.42^{***}$&& -$0.04$ & $0.18$ & -$0.25$ &&$0.03$ & $0.09$ & $0.34$ \\

Age & -$0.03$ & $0.01$ & -$3.53^{***}$&& -$0.05$ & $0.04$ & -$1.47$ &&$0.01$ & $0.02$ & $0.40$  \\

$\text{Age}^2$  & $0.0001$ \thickspace \thickspace& $0.0000$ & $4.03^{***}$ && $0.0002$\thickspace \thickspace & $0.0001$ & $3.37^{***}$ &&$0.0003$\thickspace \thickspace & $0.0001$ & $4.29^{***}$  \\

\textbf{Number of Bedrooms} \\

(Bedrooms)2 & - & - & -&& -&- &- && $0.06$ & $0.05$ & $1.41$\\ 

(Bedrooms)3 & $0.01$ & $0.05$ & $0.12$&& -$0.11$ & $0.05$ & -$2.25^{**}$ && $0.06$ & $0.06$ & $0.91$\\ 

(Bedrooms)4 & -$0.02$ & $0.05$ & -$0.41$ && -$0.12$ & $0.07$ & -$1.71^{*}$ &&$0.01$ & $0.09$ & $0.11$\\ 

(Bedrooms)5+ & -$0.05$ & $0.05$ & -$0.98$&& -$0.12$ & $0.14$ & -$0.87$ && -&-&-\\ 

Full Bathroom & $0.10$ & $0.02$ & $5.35^{***}$&& $0.02$ & $0.05$ & $0.39$ &&$0.01$ & $0.04$ & $0.15$ \\

Half Bathroom & $0.03$ & $0.02$ & $1.81^{*}$ && -$0.04$ & $0.05$ & -$0.82$ &&-$0.06$ & $0.04$ & -$1.66^{*}$\\ 

$\frac{3}{4}$ Bathroom & $0.07$ & $0.02$ & $3.98^{***}$&& -$0.05$ & $0.05$ & -$0.86$ &&$-0.001$ & $0.04$ & $-0.04$ \\

Parking & $0.01$ & $0.01$ & $0.52$&& $0.08$ & $0.04$ & $1.91^{*}$ && $0.01$ & $0.02$ & $0.59$\\ 

HOA Fees & -$0.0000$ \thickspace \thickspace \thickspace & $0.0000$ & -$0.28$&& $0.0000$ \thickspace \thickspace \thickspace & $0.0000$ & $1.00$ &&$0.0000$ \thickspace \thickspace \thickspace & $0.0000$ & $0.24$\\

\textbf{School Ranking}\\

(Nearest E.School Rank)2 & -$0.07$ & $0.02$ & -$2.87^{***}$&& -$0.24$ & $0.09$ & -$2.84^{***}$ &&-$0.06$ & $0.05$ & -$1.14$ \\

(Nearest E.School Rank)3 & -$0.20$ & $0.04$ & -$4.90^{***}$&& -$0.09$ & $0.14$ & -$0.67$ &&-$0.04$ & $0.07$ & -$0.61$ \\

(Nearest M.School Rank)2 & -$0.14$ & $0.05$ & -$3.02^{***}$ && -&- &- &&-&-&-\\

(Pool, Bath tub, Sauna, or Jacuzzi)1  \thickspace \thickspace \thickspace & $0.11$ & $0.02$ & $4.32^{***}$&& -&- &- &&-&-&- \\

(Solar Power)1 & $0.10$ & $0.05$ & $2.17^{**}$&& $0.42$ & $0.17$ & $2.39^{**}$ && - & - & - \\\bottomrule
\hline \hline
\end{tabular}
\end{adjustbox}
\end{sidewaystable}

\begin{sidewaystable} 
\setlength\tabcolsep{0pt}
\ContinuedFloat
\caption{(Continued.)}
\centering 
\ra{1.3}
\begin{adjustbox}{max width=\textwidth}
\begin{tabular}{@{}llllllllllll@{}}\toprule
\hline
\multirow{2}[3]{*}{Independent Variables} & \multicolumn{3}{c}{Single Family Housing} & \phantom{abc}& \multicolumn{3}{c}{Town-Home Housing} &
\phantom{abc} & \multicolumn{3}{c}{Condominium Housing}\\ \cmidrule{2-4} \cmidrule{6-8} \cmidrule{10-12}
& Coeff.  & Std. Error\thickspace \thickspace& T-Statistics&& Coeff.& Std. Error \thickspace \thickspace& T-Statistics && Coeff.& Std. Error \thickspace \thickspace& T-Statistics\\ \midrule
\hline\\

Drive to CBD & -$0.01$ \thickspace & $0.002$ & -$3.94^{***}$&& -$0.05$ \thickspace& $0.01$ & -$4.24^{***}$ && -$0.03$ \thickspace& $0.01$ & -$4.64^{***}$\\

Walk to E.School & -$0.001$ \thickspace& $0.001$ & -$0.77$&& $0.01$ \thickspace& $0.003$ & $2.93^{***}$  &&-$0.002$\thickspace & $0.001$ & -$1.43$ \\ 

Walk to M.school & -$0.0004$ \thickspace& $0.001$ & -$0.67$&& -$0.01$ \thickspace& $0.002$ & -$4.37^{***}$ &&-$0.003$\thickspace & $0.001$ & -$2.47^{**}$\\ 

Median Household Inc. & $0.0000$\thickspace & $0.0000$ & $0.41$&& $0.0000$ \thickspace& $0.0000$ & $0.72$  &&-$0.0000$ \thickspace& $0.0000$ & -$1.03$  \\

Neighborhood's Population& -$0.0000$ \thickspace  \thickspace& $0.0000$ & -$0.06$&& $0.0000$ \thickspace  \thickspace& $0.0000$ & $0.55$  &&-$0.0000$ \thickspace  \thickspace& $0.0000$ & -$0.71$\\

\textbf{Crime Levels} (Base: Level 3)  \thickspace \thickspace \thickspace\\

Neighborhood's Crime Level: 1 & -\thickspace & - & -&& -$0.28$ \thickspace& $0.10$ & -$2.93^{***}$ && -$0.08$ \thickspace& $0.06$ & -$1.35$\\ 

Neighborhood's Crime Level: 2 & $0.003$\thickspace & $0.02$ & $0.13$&& -$0.21$ \thickspace& $0.06$ & -$3.65^{***}$ && $0.004$ \thickspace& $0.04$ & $0.09$\\ 

\textbf{Regions} (Base: Central)\\

(North)2 & -$0.26$\thickspace & $0.04$ & -$7.10^{***}$&& -$0.14$ \thickspace& $0.15$ & -$0.99$ && -$0.04$ \thickspace& $0.07$ & -$0.60$ \\ 

(South)3 & -$0.32$\thickspace & $0.05$ & -$6.23^{***}$ && $0.13$ \thickspace& $0.20$ & $0.63$ &&$0.28$ \thickspace& $0.09$ & $2.98^{***}$\\ 

(East)4 & -$0.42$\thickspace & $0.04$ & -$9.79^{***}$&& -$0.13$ \thickspace& $0.13$ & -$0.99$  &&$0.04$ \thickspace& $0.06$ & $0.67$\\ 

(Gunbarrel)5 & -$0.51$ & $0.06$ & -$8.44^{***}$&& $1.00$ \thickspace& $0.32$ & $3.09^{***}$ && $0.42$ \thickspace& $0.16$ & $2.64^{***}$ \\ 

(Rural)6 & -$0.49$ & $0.06$ & -$8.02^{***}$&& -\thickspace&- &- &&-\thickspace&&-\\ 

Ln(Lot Area):Ln(Living Area) \thickspace \thickspace& $0.06$ & $0.02$ & $2.48^{**}$&& -& -&- &&-&-&-\\

Ln(Living Area):Age & $0.003$ & $0.001$ & $2.95^{***}$ && $0.004$ & $0.005$ & $0.78$ &&-$0.005$ & $0.003$ & -$1.68^{*}$ \\\\
\hline

Sample size & $604$ && & & $90$ &&&& $324$ \\ 

$\text{R}^2$& $0.79$ && & & $0.86$ &&&& $0.78$\\

$\text{Adj. R}^2$ & $0.78$&& & & $0.79$ &&&& $0.76$\\

F-statistic ($p-$value)& $67.01$ & $(0.00)$ &&& $13.77$ &(0.00)&&& $39.88$ & $(0.00)$ &\\
\textbf{Note}:\\ Dependent variable: Ln(Price)\\\bottomrule
\hline \hline
\end{tabular}
\end{adjustbox}
\caption*{$***,~ **$, and $*$ ~indicate $p < 0.01$, $p < 0.05$, and $p < 0.1$ respectively.\\
Base group in the single family and town-home submarkets: houses with 2 bedrooms or less in central Boulder with no solar power or pool, bath tub, sauna, or jacuzzi facilities. \\
Base group in the condominium submarket: houses with 1 bedroom or less in central Boulder with no solar power or pool, bath tub, sauna, or jacuzzi facilities \\}
\end{sidewaystable}

\begin{sidewaystable} 
\setlength\tabcolsep{0pt}
\caption{Hedonic Model Estimates - Regional Submarkets\\OLS Method.}\label{Table7}
\centering 
\ra{1.3}
\begin{adjustbox}{max width=\textwidth}
\begin{tabular}{@{}llllllllllll@{}}\toprule
\hline
\multirow{2}[3]{*}{Independent Variables} & \multicolumn{3}{c}{Central} & \phantom{abc}& \multicolumn{3}{c}{North} &
\phantom{abc} & \multicolumn{3}{c}{South}\\ \cmidrule{2-4} \cmidrule{6-8} \cmidrule{10-12}
& Coeff. & Std. Error \thickspace \thickspace& T-Statistics&& Coeff.& Std. Error \thickspace \thickspace& T-Statistics &&Coeff.& Std. Error \thickspace \thickspace& T-Statistics\\ \midrule
\hline\\

(Intercept) & $20.13$ & $3.75$ & $5.37^{***}$&& $12.91$ & $5.03$ & $2.56^{**}$  && $23.70$ & $6.43$ & $3.68^{***}$ \\ 

Ln(Lot Area) & $0.86$ & $0.19$ & $4.44^{***}$ &&-$0.12$ & $0.19$ & -$0.66$&&$0.12$ & $0.22$ & $0.54$ \\

$(\text{Ln} (\text{Lot Area}))^2$ & -$0.01$ & $0.01$ & -$0.55$ && $0.03$ & $0.01$ & $2.97^{***}$ &&$0.001$ & $0.01$ & $0.10$ \\ 

Ln(Living Area) & -$2.84$ & $1.07$ & -$2.65^{***}$ && -$0.47$ & $1.32$ &-$0.35$ &&-$3.27$ & $1.59$ & -$2.06^{**}$ \\

$(\text{Ln} (\text{Living Area}))^2$ & $0.27$ & $0.08$ & $3.45^{***}$&& $0.08$ & $0.09$ & $0.91$ &&$0.24$ & $0.10$ & $2.36^{**}$ \\

Age & -$0.01$ & $0.01$ & -$1.06$&&.  -$0.01$ & $0.02$ & -$0.52$ &&$0.02$ & $0.06$ & $0.37$ \\

$\text{Age}^2$  & $0.0000$ \thickspace \thickspace& $0.0000$ & $3.28^{***}$ && $0.0002$ \thickspace \thickspace& $0.0001$ & $2.74^{***}$ &&-$0.0003$ \thickspace \thickspace& $0.0003$ & -$0.90$  \\

\textbf{Number of Bedrooms} \\

(Bedrooms)2 & $0.01$ & $0.09$ & $0.12$ &&-$0.03$&$0.09$ & -$0.30$  && - & - & -\\ 

(Bedrooms)3 & $0.04$ & $0.10$ & $0.38$ &&   -$0.05$ & $0.11$ & -$0.41$   && $0.03$ & $0.06$ & $0.56$\\ 

(Bedrooms)4 & $0.03$ & $0.12$ & $0.29$&&   -$0.09$& $0.13$ & -$0.73$   &&$0.03$ & $0.08$ & $0.38$\\ 

(Bedrooms)5+ & -$0.06$ & $0.12$ & -$0.49$ &&.  -$0.07$ & $0.13$ & -$0.50$   && -$0.01$ & $0.08$ & -$0.14$\\ 

Full Bathroom \thickspace \thickspace \thickspace \thickspace& $0.09$ & $0.04$ & $2.59^{**}$&& $0.003$ & $0.04$ & $0.07$ &&$0.09$ & $0.03$ & $2.76^{***}$ \\

Half Bathroom & -$0.01$ & $0.04$ & -$0.18$&& -$0.07$ & $0.04$ & -$1.83^{*}$  &&$0.04$ & $0.04$ & $1.21$\\ 

$\frac{3}{4}$ Bathroom & $0.06$ & $0.04$ & $1.65$  && $0.03$ & $0.04$ & $0.73$ &&$0.07$ & $0.03$ & $2.32^{**}$\\

Parking & $0.01$ & $0.02$ & $0.42$ && $0.05$ & $0.04$ & $1.33$ && $0.01$ & $0.03$ & $0.39$ \\ 

HOA Fees &- & - & -&& $0.0000$ & $0.0000$ & $2.16^{**}$ &&- & - & -\\

(Nearest E.School Rank)2 & $0.11$ & $0.08$ & $1.38$ && -$0.16$ & $0.08$ & -$1.92^{*}$ &&-$0.10$ & $0.05$ & -$1.90^{*}$ \\ 

(Nearest E.School Rank)3 & -$0.05$ & $0.08$ & -$0.68$  &&  -$0.28$ & $0.07$ & -$4.01^{***}$&&-& - & - \\ 

(Nearest M.School Rank)2 \thickspace \thickspace \thickspace \thickspace& - & - & -  &&  - & - & -&&-$0.07$ & $0.09$ & -$0.81$  \\ \bottomrule
\hline \hline
\end{tabular}
\end{adjustbox}
\end{sidewaystable} 

\begin{sidewaystable} 
\setlength\tabcolsep{0pt}
\ContinuedFloat
\caption{Continued.}\label{Table7}
\centering 
\ra{1.3}
\begin{adjustbox}{max width=\textwidth}
\begin{tabular}{@{}llllllllllll@{}}\toprule
\hline
\multirow{2}[3]{*}{Independent Variables} & \multicolumn{3}{c}{Central} & \phantom{abc}& \multicolumn{3}{c}{North} &
\phantom{abc} & \multicolumn{3}{c}{South}\\ \cmidrule{2-4} \cmidrule{6-8} \cmidrule{10-12}
& Coeff. & Std. Error \thickspace \thickspace& T-Statistics&& Coeff.& Std. Error \thickspace \thickspace& T-Statistics &&Coeff.& Std. Error \thickspace \thickspace& T-Statistics\\ \midrule
\hline\\

Drive to CBD & -$0.06$ & $0.01$ & -$4.84^{***}$&& -$0.05$ & $0.02$ & -$2.81^{***}$ && $0.01$ & $0.01$ & $0.79$\\

Walk to E.School & $0.01$ & $0.003$ & $2.61^{***}$&& -$0.01$ & $0.004$ & -$1.75^{*}$ &&-$0.01$ & $0.003$ & -$1.65$ \\ 

Walk to M.school & -$0.001$ & $0.001$ & -$1.06$&& $0.01$ & $0.01$ & $1.60$ &&$0.01$ & $0.003$ & $2.62^{**}$\\ 

Median Household Inc. & $0.0000$ & $0.0000$ & $1.03$ &&  $0.0000$ & $0.0000$ & $1.04$  && - & - & - \\

Neighborhood's Population & $0.0000$ \thickspace \thickspace& $0.0000$ & $0.66$ && -$0.0000$ \thickspace \thickspace& $0.0000$ & -$1.65$ &&$0.0000$ \thickspace \thickspace & $0.0000$ & $0.29$\\

\textbf{Crime Levels} (Base: Level 3) \thickspace \thickspace \\

Neighborhood's Crime Level: 1 & - & - & -&& -$0.02$ & $0.07$ & -$0.28$ && - & - & -\\ 

Neighborhood's Crime Level: 2 & -$0.04$ & $0.05$ & -$0.76$&& $0.03$ & $0.04$ & $0.79$ && -$0.05$ & $0.04$ & -$1.28$\\ 
 
\textbf{Property Types} (Base: Condominium) \thickspace \thickspace\\

(Town-Home)2 & -$1.31$ & $0.62$ & -$2.09^{**}$&&$1.21$ & $0.65$ & $1.87^{*}$ && -$0.55$ & $0.65$ & -$0.84$ \\ 

(Single Family)3 & -$1.24$ & $0.66$ & -$1.89^{*}$ && $1.44$ & $0.66$ & $2.19^{**}$ &&-$0.50$ & $0.69$ & -$0.72$\\ 

Ln(Lot Area):Ln(Living Area) & -$0.09$ & $0.02$ & -$4.75^{***}$&&-$0.03$& $0.01$&-$2.55^{**}$&&-$0.004$ & $0.02$ & -$0.17$ \\

Ln(Living Area):Age & -$0.0001$ & $0.001$ & -$0.11$ &&  -$0.001$ & $0.003$ & -$0.28$ &&-$0.0004$ & $0.01$ & -$0.06$ \\\\
\hline

Sample size & $230$&& & &$238$ &&&& $143$ \\ 

$\text{R}^2$& $0.9$ && & & $0.88$ && & &$0.88$\\

$\text{Adj. R}^2$& $0.88$&& & &$0.87$ && & & $0.86$\\

F-statistic ($p-$value)&$67.31$&$(0.00)$&&&$55.6$&(0.00)&&&$37.33$&(0.00)&\\
Dependent variable: Ln(Price)\\\bottomrule
\hline \hline
\end{tabular}
\end{adjustbox}
\caption*{$***,~ **$, and $*$ ~indicate $p < 0.01$, $p < 0.05$, and $p < 0.1$ respectively.\\
Base group in the central, north, east Gunbarrel submarkets: condominiums with 1 bedroom or less with no solar power or pool, bath tub, sauna, or jacuzzi facilities. \\
Base group in south and rural submarkets: condominiums with 2 bedroom or less with no solar power or pool, bath tub, sauna, or jacuzzi facilities.\\}
\end{sidewaystable} 

\begin{sidewaystable} 
\setlength\tabcolsep{0pt}
\ContinuedFloat
\caption{Continued.}\label{Table7}
\centering 
\ra{1.3}
\begin{adjustbox}{max width=\textwidth}
\begin{tabular}{@{}llllllllllll@{}}\toprule
\hline
\multirow{2}[3]{*}{Independent Variables} & \multicolumn{3}{c}{East} & \phantom{abc}& \multicolumn{3}{c}{Gunbarrel} &
\phantom{abc} & \multicolumn{3}{c}{Rural}\\ \cmidrule{2-4} \cmidrule{6-8} \cmidrule{10-12}
& Coeff. & Std. Error \thickspace \thickspace& T-Statistics&& Coeff.& Std. Error\thickspace \thickspace& T-Statistics &&Coeff.& Std. Error \thickspace \thickspace & T-Statistics\\ \midrule
\hline\\

(Intercept) & $13.26$ & $4.18$ & $3.17^{***}$ && $23.58$ & $3.54$ & $6.67^{***}$  && $28.84$ & $10.46$ & $2.76^{***}$ \\ 

Ln(Lot Area) & $0.46$ & $0.15$ & $3.01^{***}$  && - & - & -  &&$0.02$ & $0.79$ & $0.03$  \\

$(\text{Ln} (\text{Lot Area}))^2$ & -$0.001$ & $0.01$ & -$0.20$ &&  - & - & -  && $0.004$ & $0.04$ & $0.08$ \\ 

Ln(Living Area) & -$0.80$ & $1.17$ & -$0.68$ && -$2.66$ & $0.85$ & -$3.11^{***}$ &&-$3.73$ & $2.46$ & -$1.51$ \\

$(\text{Ln} (\text{Living Area}))^2$ & $0.10$ & $0.08$ & $1.28$ && $0.17$ & $0.05$ & $3.26^{***}$  && $0.24$ & $0.16$ & $1.50$ \\

Age & -$0.02$ & $0.02$ & -$1.05$&&  -$0.12$ & $0.03$ & -$3.86^{***}$ &&-$0.17$ & $0.06$ & -$2.93^{***}$ \\

$\text{Age}^2$  & $0.0002$ & $0.0001^{***}$ & $3.34$  && $0.0002$ & $0.0002$ & $0.96$ &&$0.001$ & $0.0002$ & $2.87^{***}$ \\

\textbf{Number of Bedrooms} \\

(Bedrooms)2 & $0.06$ & $0.06$ & $1.09$  && $0.11$ & $0.06$ & $1.92^{*}$ && - & - & -\\

(Bedrooms)3 & $0.02$ & $0.08$ & $0.30$ &&   $0.07$ & $0.12$ & $0.55$   && -$0.01$ & $0.13$ & -$0.09$\\ 

(Bedrooms)4 & $0.01$ & $0.09$ & $0.07$&&   $0.04$ & $0.12$ & $0.36$  &&-$0.03$ & $0.15$ & -$0.19$\\ 

(Bedrooms)5+ & $0.11$ & $0.11$ & $1.05$  && -$0.04$ & $0.13$ & -$0.29$  && -$0.01$ & $0.20$ & -$0.07$\\ 

Full Bathroom \thickspace \thickspace \thickspace& $0.04$ & $0.04$ & $1.19$ && $0.12$ & $0.04$ & $3.12^{***}$ &&$0.05$ & $0.07$ & $0.68$  \\

Half Bathroom & $0.01$ & $0.03$ & $0.24$ && $0.11$ & $0.04$ & $3.07^{***}$ &&$0.07$ & $0.08$ & $0.79$\\ 

$\frac{3}{4}$ Bathroom & $0.07$ & $0.03$ & $2.14^{**}$  && $0.08$ & $0.04$ & $2.19^{**}$ &&-$0.01$ & $0.07$ & -$0.15$\\

Parking & $0.02$ & $0.02$ & $0.80$&& $0.06$ & $0.03$ & $2.33^{**}$ && -$0.02$ & $0.04$ & -$0.56$ \\ 

HOA Fees &$0.0000$ \thickspace \thickspace \thickspace& $0.0000$ & $1.98^{**}$ && -$0.0000$ \thickspace \thickspace \thickspace& $0.0000$ & -$1.17$ &&$0.0001$ \thickspace \thickspace \thickspace& $0.0001$ & $0.77$\\

(Nearest E.School Rank)2 & $0.03$ & $0.05$ & $0.63$ && - & - & - &&-$0.10$ & $0.11$ & -$0.91$  \\ 

(Nearest E.School Rank)3 & -$0.05$ & $0.08$ & -$0.68$  &&  -& - & -&&-& - & - \\ 

(Nearest M.School Rank)2 \thickspace \thickspace \thickspace \thickspace& - & - & -  &&  - & - & -&&$0.08$ & $0.18$ & $0.44$   \\\bottomrule
\hline \hline
\end{tabular}
\end{adjustbox}
\end{sidewaystable}

\begin{sidewaystable} 
\setlength\tabcolsep{0pt}
\ContinuedFloat
\caption{Continued.}\label{Table7}
\captionsetup{singlelinecheck = false}
\centering 
\ra{1.3}
\begin{adjustbox}{max width=\textwidth}
\begin{tabular}{@{}llllllllllll@{}}\toprule
\hline
\multirow{2}[3]{*}{Independent Variables} & \multicolumn{3}{c}{East} & \phantom{abc}& \multicolumn{3}{c}{Gunbarrel} &
\phantom{abc} & \multicolumn{3}{c}{Rural}\\ \cmidrule{2-4} \cmidrule{6-8} \cmidrule{10-12}
& Coeff. & Std. Error \thickspace& T-Statistics&& Coeff.& Std. Error \thickspace& T-Statistics &&Coeff.& Std. Error \thickspace& T-Statistics\\ \midrule
\hline\\
 (Pool, Bath tub, Sauna, or Jacuzzi)1 & - & - & -&& -&- &- &&  $0.13$ & $0.10$ & $1.33$ \\

(Solar Power)1 & - & - & - && - & - & - && $0.14$ & $0.18$ & $0.78$ \\

Drive to CBD & $0.02$ & $0.01$ & $1.72^{*}$ && -$0.01$ & $0.01$ & -$0.81$ && -$0.01$ & $0.003$ & -$2.57^{**}$\\

Walk to E.School & $0.003$ & $0.002$ & $1.50$ && - & - & - &&$0.001$ & $0.002$ & $0.64$  \\ 

Walk to M.school & $0.0000$ & $0.002$ & $0.003$  && -$0.003$ & $0.001$ & -$1.88^{*}$ &&$0.01$ & $0.003$ & $2.62^{**}$\\ 

Median Household Inc. & -$0.0000$ & $0.0000$ & -$2.47^{**}$ &&  -$0.0000$ & $0.0000$ & -$0.38$  && $0.0000$ & $0.0000$ & $0.10$ \\

Neighborhood's Population & $0.0000$ \thickspace \thickspace \thickspace& $0.0000$ & $4.43^{***}$ && $0.0001$ \thickspace \thickspace \thickspace& $0.0000$ & $2.65^{***}$ &&-$0.0000$ \thickspace \thickspace \thickspace& $0.0000$ & -$0.74$\\

\textbf{Crime Levels} (Base: Level 3) \thickspace \thickspace\\

Neighborhood's Crime Levels: 1 &$0.10$ & $0.07$ & $1.39$ && - & - & - && - & - & -\\ 

Neighborhood's Crime Levels: 2 & -$0.0000$ & $0.04$ & -$0.0004$ && -$0.04$ & $0.05$ & -$0.76$ && $0.09$ & $0.12$ & $0.74$\\ 
 
\textbf{Property Types} (Base: Condominium) \thickspace \thickspace \\

(Town-Home)2 & -$1.10$ & $0.34$ & -$3.18^{***}$ && $0.11$ & $0.14$ & $0.80$  && -$0.55$ & $0.65$ & -$0.84$ \\ 

(Single Family)3 & -$1.08$ & $0.36$ & -$3.00^{***}$ && $0.14$ & $0.15$ & $0.95$ &&-$0.50$ & $0.69$ & -$0.72$\\ 

Ln(Lot Area):Ln(Living Area) & -$0.04$ & $0.02$ & -$2.24^{**}$  &&-& -&- &&-$0.01$ & $0.12$ & -$0.07$ \\

Ln(Living Area):Age & $0.0000$ & $0.002$ & $0.02$ &&  $0.01$ & $0.003$ & $4.20^{***}$ &&$0.02$ & $0.01$ & $2.34^{**}$ \\\\
\hline

Sample size & $200$&& & &$125$ &&&& $82$ \\ 

$\text{R}^2$& $0.91$ && & & $0.93$ && & &$0.81$\\

$\text{Adj. R}^2$& $0.89$&& & &$0.91$ && & & $0.73$\\

F-statistic ($p-$value)&$63.19$&$(0.00)$&&&$62.57$&($0.00$)&&&$9.57$&($0.00$)&\\
Dependent variable.: Ln(Price)\\\bottomrule
\hline \hline
\end{tabular}
\end{adjustbox}
\caption*{$***,~ **$, and $*$ ~indicate $p < 0.01$, $p < 0.05$, and $p < 0.1$ respectively.\\}
\end{sidewaystable}

\begin{table}[ht]
\caption{Breusch-Pagan Results.}\label{Table8}
\captionsetup{singlelinecheck = false}
\centering
\begin{adjustbox}{max width=\textwidth}
\begin{tabular}{llllllcccccccllllllll}
\hline\hline\\
Market level &&&&&&&&&&&&&&&&&&&& $\emph{P}$-value \\ [0.5ex] 
\hline \hline
\\
Citywide  &&&&&&&&&&&&&&&&&&&&$0.0000$\\
\\
Single Family &&&&&&&&&&&&&&&&&&&&$0.0000$\\
\\
Town-Home   &&&&&&&&&&&&&&&&&&&& $0.03^{***}$\\
\\
Condominium &&&&&&&&&&&&&&&&&&&&$0.0000$\\
\\
Central &&&&&&&&&&&&&&&&&&&&$0.09^{***}$\\
\\
North &&&&&&&&&&&&&&&&&&&&$0.005$\\
\\
South&&&&&&&&&&&&&&&&&&&&$0.03^{***}$\\
\\
East&&&&&&&&&&&&&&&&&&&&$0.0001$\\
\\
Gunbarrel&&&&&&&&&&&&&&&&&&&&$0.03^{***}$\\
\\
Rural&&&&&&&&&&&&&&&&&&&&$0.2^{***}$\\\\[1ex]
\\
\hline \hline
\end{tabular}
\end{adjustbox}
\label{table:nonlin}
\caption*{\textbf{Note}:  $***$~ indicates significance at $1\%$. \thickspace \thickspace \\  \thickspace \thickspace}
\end{table}

\begin{table}[ht]
\caption{F-test Results for Property-Type Submarkets.}\label{Table9}
\captionsetup{singlelinecheck = false}
\centering
\begin{adjustbox}{max width=\textwidth}
\begin{tabular}{lccccccccccccl}
\hline\hline\\
Submarkets &&&&&&&&&&&&& Chow \\ [0.5ex] 
\hline \hline
\\
Single Family with Condominium&&&&&&&&&&&&&$4.79^{***}$\\
\\
Single Family with Town-Home &&&&&&&&&&&&&$2.38^{***}$\\
\\
Town-Home with Condominium &&&&&&&&&&&&& $1.25$\\\\[1ex]

\hline \hline
\end{tabular}
\end{adjustbox}
\label{table:nonlin}
\caption*{\textbf{Note}:  $***$~ indicates significance at $1\%$. \thickspace \thickspace \\  \thickspace \thickspace}
\end{table}

\begin{table}[ht]
\caption{F-test Results for Spatial Submarkets.}\label{Table10}
\captionsetup{singlelinecheck = false}
\centering
\begin{adjustbox}{max width=\textwidth}
\begin{tabular}{lcccccccccccccccclll}
\hline\hline\\
Submarkets \thickspace \thickspace \thickspace&&&&&&&&&&&&&&&&& \thickspace Chow \\ [0.5ex] 
\hline \hline
\\
Central with North&&&&&&&&&&&&&&&&&$1.7^{**}$\\
\\
Central with South &&&&&&&&&&&&&&&&&$2.37^{***}$\\
\\
Central with East &&&&&&&&&&&&&&&&&$3.16^{***}$ \\ 
\\
Central with Gunbarrel &&&&&&&&&&&&&&&&&$2.45^{***}$ \\ 
\\
Central with Rural &&&&&&&&&&&&&&&&&$1.95^{***}$ \\ 
\\
North with South &&&&&&&&&&&&&&&&&$0.95$ \\ 
\\
North with East &&&&&&&&&&&&&&&&&$1.96^{***}$\\ 
\\
North with Gunbarrel &&&&&&&&&&&&&&&&&$1.8^{**}$ \\ 
\\
North with Rural &&&&&&&&&&&&&&&&&$1.37$ \\
\\
South with East &&&&&&&&&&&&&&&&&$1.89^{***}$ \\ 
\\
South with Gunbarrel &&&&&&&&&&&&&&&&&$0.89$ \\ 
\\
South with Rural &&&&&&&&&&&&&&&&&$1.42^{*}$ \\ 
\\
East with Gunbarrel &&&&&&&&&&&&&&&&&$1.96^{***}$ \\
\\
East with Rural &&&&&&&&&&&&&&&&&$1.75^{**}$ \\ 
\\
Gunbarrel with Rural &&&&&&&&&&&&&&&&&$1.05$ \\ 
\\[1ex]
\hline \hline
\end{tabular}
\end{adjustbox}
\label{table:nonlin}
\caption*{\textbf{Note}: $***,~ **$, and $*$ ~indicate significance at $1\%$, $5\%$ and $10\%$ respectively.\\ }
\end{table}

\begin{sidewaystable} 
\setlength\tabcolsep{0pt}
\caption{Hedonic Model Estimates - Nested Locational and Structural Submarkets -- OLS Method.}\label{Table12}
\centering 
\ra{1.3}
\begin{adjustbox}{max width=\textwidth}
\begin{tabular}{@{}llllllllllll@{}}\toprule
\hline
\multirow{2}[3]{*}{Independent Variables} & \multicolumn{3}{c}{Central-Single Family} & \phantom{abc}& \multicolumn{3}{c}{Central-Condo-Town-Home} &
\phantom{abc} & \multicolumn{3}{c}{NSGR-Single Family}\\ \cmidrule{2-4} \cmidrule{6-8} \cmidrule{10-12}
& Coeff. \thickspace & Std. Error \thickspace& T-Statistics&& Coeff. \thickspace& Std. Error \thickspace& T-Statistics &&Coeff.\thickspace \thickspace& Std. Error \thickspace& T-Statistics\\ \midrule
\hline\\

(Intercept) & $15.67$ & $7.36$ & $2.13^{**}$ && $21.51$ & $4.70$ & $4.58^{***}$  &&  $28.73$ & $3.30$ & $8.71^{***}$  \\ 

Ln(Lot Area) & $1.46$ & $0.76$ & $1.93^{*}$ && - & - & - && $0.15$ & $0.19$ & $0.76$  \\

$(\text{Ln} (\text{Lot Area}))^2$ & -$0.14$ & $0.04$ & -$3.25^{***}$&& -& - & - &&-$0.03$ & $0.01$ & -$3.20^{***}$\\ 

Ln(Living Area) & -$2.74$ & $1.52$ & -$1.80^{*}$ && -$3.12$ & $1.38$ & -$2.27^{**}$  && -$3.94$ & $0.77$ & -$5.10^{***}$  \\

$(\text{Ln} (\text{Living Area}))^2$ & $0.10$ & $0.10$ & $1.00$ && $0.28$ & $0.10$ & $2.79^{***}$  &&$0.22$ & $0.05$ & $4.37^{***}$\\

Age & -$0.02$ & $0.01$ & -$1.11$ && -$0.002$ & $0.01$ & -$0.12$ &&-$0.10$ & $0.02$ & -$5.20^{***}$  \\

$\text{Age}^2$  & $0.0000$ & $0.0000$ & $1.54$ &&  $0.0001$ \thickspace& $0.0000$ & $4.31^{***}$ &&$0.0003$ \thickspace \thickspace& $0.0001$ & $5.28^{***}$\\

\textbf{Number of Bedrooms} \\

(Bedrooms)2 & - & - & -&&$0.05$ & $0.08$ & $0.65$&& - & -&-\\

(Bedrooms)3 & $0.06$ & $0.08$ & $0.79$ &&  $0.15$ & $0.12$ & $1.31$&&  -$0.03$ & $0.07$ & -$0.47$\\ 

(Bedrooms)4 & $0.06$ & $0.09$ & $0.64$ &&  $0.01$ & $0.15$ & $0.10$&&-$0.10$ & $0.07$ & -$1.33$ \\ 

(Bedrooms)5+ & -$0.04$ & $0.10$ & -$0.37$ && - & - &- && -$0.10$ & $0.08$ & -$1.36$\\ 

Full Bathroom \thickspace \thickspace \thickspace& $0.12$ & $0.04$ & $3.09^{***}$&& $0.10$ & $0.07$ & $1.44$  &&$0.08$ & $0.02$ & $3.09^{***}$ \\

Half Bathroom & -$0.01$ & $0.05$ & -$0.12$ && $0.02$ & $0.06$ & $0.39$ && $0.03$ & $0.02$ & $1.27$ \\ 

$\frac{3}{4}$ Bathroom & $0.06$ & $0.04$ & $1.51$  && $0.09$ & $0.06$ & $1.49$  &&$0.07$ & $0.02$ & $2.76^{***}$\\

Parking & -$0.005$ & $0.02$ & -$0.20$ && $0.03$ & $0.04$ & $0.93$  &&  $0.03$ & $0.02$ & $1.39$\\ 

HOA Fees &- & - & -&& - & - & -&&- & - & -\\

(Nearest E.School Rank)2 & $0.21$ & $0.09$ & $2.33^{**}$ &&   $0.29$ & $0.29$ & $0.98$  &&-$0.25$ & $0.03$ & -$10.02^{***}$  \\ 

(Nearest E.School Rank)3 & -$0.10$ & $0.08$ & -$1.18$  &&  $0.15$ & $0.29$ & $0.54$ &&-$0.28$ & $0.05$ & -$5.20^{***}$ \\ 

(Nearest M.School Rank)2 \thickspace \thickspace \thickspace \thickspace& - & - & -  &&  - & - & -&&- & - & -  \\ \bottomrule
\hline \hline
\end{tabular}
\end{adjustbox}
\end{sidewaystable}

\begin{sidewaystable} 
\setlength\tabcolsep{0pt}
\ContinuedFloat
\caption{Continued.}\label{Table12}
\centering 
\ra{1.3}
\begin{adjustbox}{max width=\textwidth}
\begin{tabular}{@{}llllllllllll@{}}\toprule
\hline
\multirow{2}[3]{*}{Independent Variables} & \multicolumn{3}{c}{Central-Single Family} & \phantom{abc}& \multicolumn{3}{c}{Central-Condo-Town-Home} &
\phantom{abc} & \multicolumn{3}{c}{NSGR-Single Family}\\ \cmidrule{2-4} \cmidrule{6-8} \cmidrule{10-12}
& Coeff. & Std. Error \thickspace& T-Statistics&& Coeff.& Std. Error& \thickspace T-Statistics &&Coeff.& Std. Error& \thickspace T-Statistics\\ \midrule
\hline\\

(Pool, Bath tub, Sauna, or Jacuzzi)1 \thickspace \thickspace \thickspace& $0.06$ & $0.06$ & $0.97$ && - & - & -&& $0.11$ & $0.03$ & $3.64^{***}$  \\ 

(Solar Power)1 & -$0.02$ & $0.18$ & -$0.09$ && - & - & - && $0.09$ & $0.06$ & $1.51$ \\
Drive to CBD  & -$0.04$ & $0.02$ & -$2.34^{**}$  && -$0.06$ & $0.02$ & -$3.17^{***}$ &&  -$0.01$ & $0.002$ & -$6.02^{***}$\\

Walk to E.School & $0.003$ & $0.004$ & $0.63$ && $0.01$ & $0.004$ & $3.57^{***}$ &&-$0.003$ & $0.001$ & -$3.51^{***}$ \\ 

Walk to M.school & -$0.0003$ & $0.002$ & -$0.19$ && -$0.01$ & $0.001$ & -$3.96^{***}$&&- &- & -\\ 

Median Household Inc. & $0.0000$ & $0.0000$ & $1.49$ &&  $0.0000$ & $0.0000$ & $0.11$   && -$0.0000$ & $0.0000$ & -$2.77^{***}$  \\

Neighborhood's Population & $0.0000$ \thickspace \thickspace \thickspace& $0.0000$ & $1.68^{*}$ && -$0.0000$ \thickspace \thickspace \thickspace& $0.0000$ & -$0.17$ &&$0.0000$ \thickspace \thickspace \thickspace& $0.0000$ & $1.36$  \\

\textbf{Crime Levels} (Base: Level 3) \\

Neighborhood's Crime Level: 1 & - & - & -&& -& - & - &&-$0.19$ & $0.21$ & -$0.91$ \\ 

Neighborhood's Crime Level: 2 & $0.04$ & $0.05$ & $0.79$ && -$0.11$ & $0.13$ & -$0.83$  && -$0.02$ & $0.03$ & -$0.89$ \\ 

Ln(Lot Area):Ln(Living Area) & $0.17$ & $0.12$ & $1.41$ && - & -& - &&$0.06$ & $0.03$ & $2.25^{**}$\\

Ln(Living Area):Age & $0.001$ & $0.002$ & $0.73$ && -$0.002$ & $0.002$ & -$1.08$ &&$0.01$ & $0.002$ & $4.46^{***}$ \\\\
\hline

Sample size & $144$&& & &$86$ &&&& $373$ \\ 

$\text{R}^2$& $0.83$ && & & $0.94$ && & &$0.75$\\

$\text{Adj. R}^2$& $0.79$&& & &$0.92$ && & & $0.74$\\

F-statistic ($p-$value)&$23.08$&$(0.00)$&&&  $47.18$&(0.00)&&&    $42.42$&(0.00)&\\\bottomrule
\hline \hline
\end{tabular}
\end{adjustbox}
\end{sidewaystable}

\begin{sidewaystable*} 
\setlength\tabcolsep{0pt}
\ContinuedFloat
\caption{Continued.}\label{Table12}
\centering 
\ra{1.3}
\begin{adjustbox}{max width=\textwidth}
\begin{tabular}{@{}llllllllllll@{}}\toprule
\hline
\multirow{2}[3]{*}{Independent Variables} & \multicolumn{3}{c}{NSGR-Condo-Town-Home} & \phantom{abc}& \multicolumn{3}{c}{East-Single Family} &
\phantom{abc} & \multicolumn{3}{c}{East-Condo-Town-Home}\\ \cmidrule{2-4} \cmidrule{6-8} \cmidrule{10-12}
& Coeff. \thickspace& Std. Error \thickspace& T-Statistics&& Coeff.& Std. Error \thickspace& T-Statistics &&Coeff.& Std. Error \thickspace & T-Statistics\\ \midrule
\hline\\

(Intercept) & $4.17$ & $4.21$ & $0.99$  &&  $6.78$ & $8.42$ & $0.80$  &&  -$0.18$ & $4.04$ & -$0.04$ \\ 

Ln(Lot Area) & - & -& - && $0.11$ & $0.40$ & $0.28$ && - & - & -  \\

$(\text{Ln} (\text{Lot Area}))^2$ & -& - & - && $0.002$ & $0.01$ & $0.25$ &&- & - & -\\

Ln(Living Area) & $1.97$ & $1.10$ & $1.79^{*}$  && $0.41$ & $2.07$ & $0.20$ &&$2.90$ & $1.10$ & $2.65^{***}$ \\

$(\text{Ln} (\text{Living Area}))^2$ & -$0.09$ & $0.07$ & -$1.22$ && $0.03$ & $0.13$ & $0.23$&&-$0.15$ & $0.07$ & -$2.02^{**}$ \\

Age & $0.01$ & $0.02$ & $0.52$ && $0.08$ & $0.04$ & $2.25^{**}$ &&$0.03$ & $0.02$ & $1.26$  \\

$\text{Age}^2$  & $0.0000$ & $0.0001$ & $0.09$ &&  -$0.0000$ \thickspace \thickspace \thickspace& $0.0001$ & -$0.09$  &&$0.0002$ & $0.0001$ & $3.44^{***}$  \\

\textbf{Number of Bedrooms} \\

(Bedrooms)2 & $0.02$ & $0.06$ & $0.27$ && -&- & -  && $0.10$ & $0.05$ & $2.16^{**}$\\

(Bedrooms)3 & $0.01$ & $0.07$ & $0.19$  &&   - & - & -  &&  $0.17$ & $0.06$ & $2.73^{***}$\\ 

(Bedrooms)4 & -$0.05$ & $0.09$ & -$0.55$  &&  -$0.06$ & $0.06$ & -$1.01$  && $0.08$ & $0.09$ & $0.86$\\

(Bedrooms)5+ & - & - & - && -$0.01$ & $0.08$ & -$0.17$  && - & - & -\\ 

Full Bathroom & -$0.02$ & $0.04$ & -$0.48$&& $0.13$ & $0.07$ & $2.03^{**}$  && -$0.01$ & $0.04$ & -$0.36$\\

Half Bathroom & -$0.09$ & $0.04$ & -$2.51^{**}$ && $0.04$ & $0.04$ & $0.88$ &&
$0.02$ & $0.04$ & $0.45$\\ 

$\frac{3}{4}$ Bathroom & -$0.01$ & $0.04$ & -$0.20$  && $0.17$ & $0.05$ & $3.20^{***}$  &&$0.01$ & $0.04$ & $0.17$\\

Parking & $0.04$ & $0.03$ & $1.51$  && $0.01$ & $0.05$ & $0.27$ &&  $0.01$ & $0.02$ & $0.82$ \\ 

HOA Fees &- & - & -&& $0.0001$ & $0.0001$ & $1.46$ &&$0.0000$ \thickspace \thickspace \thickspace & $0.0000$ & $2.30^{**}$\\

(Nearest E.School Rank)2 & -$0.29$ & $0.04$ & -$6.78^{***}$ &&  $0.04$ & $0.07$ & $0.60$ &&- & - & -  \\

(Nearest E.School Rank)3 & -$0.33$ & $0.06$ & -$5.82^{***}$  &&  - & - & -&&-& - & - \\ 

(Nearest M.School Rank)2\thickspace \thickspace \thickspace \thickspace & - & - & -  &&  - & - & -&&- & - & -  \\ \bottomrule
\hline \hline
\end{tabular}
\end{adjustbox}
\end{sidewaystable*}

\begin{sidewaystable*} 
\setlength\tabcolsep{0pt}
\ContinuedFloat
\caption{Continued.}\label{Table12}
\captionsetup{singlelinecheck = false}
\centering 
\ra{1.3}
\begin{adjustbox}{max width=\textwidth}
\begin{tabular}{@{}llllllllllll@{}}\toprule
\hline
\multirow{2}[3]{*}{Independent Variables} & \multicolumn{3}{c}{NSGR-Condo-Town-Home} & \phantom{abc}& \multicolumn{3}{c}{East-Single Family} &
\phantom{abc} & \multicolumn{3}{c}{East-Condo-Town-Home}\\ \cmidrule{2-4} \cmidrule{6-8} \cmidrule{10-12}
& Coeff. & Std. Error \thickspace& T-Statistics&& Coeff.& Std. Error \thickspace& T-Statistics &&Coeff.& Std. Error \thickspace& T-Statistics\\ \midrule
\hline\\
(Pool, Bath tub, Sauna, or Jacuzzi)1 \thickspace \thickspace \thickspace& -& - & - && $0.09$ & $0.09$ & $1.05$ &&  - & - & -  \\ 

(Solar Power)1 & - & - & - && $0.06$ & $0.09$ & $0.63$  && - & - & -  \\ 

Drive to CBD  & -$0.01$ & $0.003$ & -$3.79^{***}$  &&  $0.02$ & $0.02$ & $1.38$  &&  $0.01$ & $0.01$ & $1.75^{*}$\\

Walk to E.School & -$0.002$ & $0.001$ & -$2.04^{**}$&& -$0.01$ & $0.004$ & -$1.81^{*}$ &&$0.01$ & $0.002$ & $4.30^{***}$   \\ 

Walk to M.school & - & - & - && $0.001$ & $0.003$ & $0.35$ &&- & - & -\\ 

Median Household Inc. &-$0.0000$ \thickspace \thickspace \thickspace& $0.0000$ & -$1.16$&&  -$0.0000$\thickspace \thickspace \thickspace & $0.0000$ & -$0.56$ && $0.0000$ \thickspace \thickspace \thickspace& $0.0000$ & $0.52$ \\

Neighborhood's Population & -$0.0000$ & $0.0000$ & -$1.50$ && $0.0000$ & $0.0000$ & $0.08$ &&$0.0000$ \thickspace \thickspace \thickspace & $0.0000$ & $4.62^{***}$\\

\textbf{Crime Levels} (Base: Level 3) \thickspace \thickspace\\

Neighborhoods' Crime Level: 1 & -$0.08$ & $0.07$ & -$1.24$&& - & - & - &&$0.07$ & $0.06$ & $1.16$ \\ 

Neighborhoods' Crime Level: 2 & -$0.05$ & $0.04$ & -$1.20$ &&   $0.03$ & $0.06$ & $0.57$  && -$0.11$ & $0.06$ & -$1.97^{*}$ \\ 

Ln(Lot Area):Ln(Living Area) & - & - & - && -$0.01$ & $0.06$ & -$0.10$ &&- & - & - \\

Ln(Living Area):Age & -$0.002$ & $0.003$ & -$0.71$  &&  -$0.01$ & $0.005$ & -$1.97^{*}$ && -$0.01$ & $0.003$ & -$2.15^{**}$ \\\\
\hline

Sample size & $215$&& & &$87$ &&&& $113$ \\ 

$\text{R}^2$& $0.78$ && & & $0.85$ && & &$0.9$\\

$\text{Adj. R}^2$& $0.76$&& & &$0.79$ && & & $0.89$\\

F-statistic ($p-$value)&$34.07$&$(0.00)$&&&  $14.19$&($0.00$)&&&    $46.55$&($0.00$)&\\
Dependent variable: Ln(Price)\\\bottomrule
\hline \hline
\end{tabular}
\end{adjustbox}
\caption*{\textbf{Note}: $***,~ **$, and $*$ ~indicate $p < 0.01$, $p < 0.05$, and $p < 0.1$ respectively.\\}
\end{sidewaystable*}

\begin{table}[!hp]
\caption{F-Test Results for Nested Submarkets.}\label{Table13}
\centering
\begin{adjustbox}{max width=\textwidth}
\begin{tabular}{lllllll}
\hline\hline
\\
Submarkets &&&&& Chow \\ [0.99ex] 
\hline \hline
\\
Central-Single Family with Central-Condo-Town-Home&&&&&$4.03^{***}$\\
\\
Central-Single Family with NSGR-Single Family&&&&&$3.33^{***}$\\
\\
Central-Single Family with NSGR-Condo-Town-Home&&&&&$0.76$\\
\\
Central-Single Family with East-Single Family&&&&&$3.32^{***}$\\
\\
Central-Single Family with East-Condo-Town-Home&&&&&$2.06^{***}$\\
\\
Central-Condo-Town-Home with NSGR-Single Family&&&&&$5.79^{***}$\\
\\
Central-Condo-Town-Home with NSGR-Condo-Town-Home&&&&&$4.55^{***}$\\
\\
Central-Condo-Town-Home with East-Single Family&&&&&$8.87^{***}$\\
\\
Central-Condo-Town-Home with East-Condo-Town-Home&&&&&$7.11^{***}$\\
\\

NSGR-Single Family with NSGR-Condo-Town-Home &&&&&$1.39^{*}$\\
\\
NSGR-Single Family with East-Single Family &&&&&$6.63^{***}$\\
\\
NSGR-Single Family with East-Condo-Town-Home &&&&&$3.34^{***}$\\
\\
NSGR-Condo-Town-Home with East-Single Family &&&&&$2.66^{***}$ \\
\\
NSGR-Condo-Town-Home with East-Condo-Town-Home &&&&&0.68 \\

\\
East-Single Family with East-Condo-Town-Home &&&&&$3.16^{***}$  \\ [1ex]
\hline \hline
\end{tabular}
\end{adjustbox}
\label{table:nonlin}
\caption*{\textbf{Note}: $***,~ **$, and $*$ ~indicate significance at $1\%$, $5\%$ and $10\%$ respectively.\\ }
\end{table}

\end{appendix}

\end{document}